\documentclass[conference]{IEEEtran}
\IEEEoverridecommandlockouts
\usepackage{cite}

\usepackage{algorithmic}
\usepackage{graphicx}
\usepackage{textcomp}
\usepackage{xcolor}
\usepackage{threeparttable}
\usepackage{tabularx,adjustbox}
\usepackage{longfbox}
\usepackage{subfigure}
\usepackage{multirow}
\usepackage{capt-of}
\usepackage[normalem]{ulem}
\usepackage{amsmath,amsfonts,amssymb,mathtools,mathrsfs}

\usepackage{epsfig,amstext,xspace}

\usepackage{amsthm}

\usepackage[sans]{dsfont}
\usepackage{stmaryrd}
\usepackage{pifont}
\usepackage{wasysym}
\usepackage{hhline}

\usepackage{array}
\usepackage{url}

\usepackage{enumitem}

\def\BibTeX{{\rm B\kern-.05em{\sc i\kern-.025em b}\kern-.08em
    T\kern-.1667em\lower.7ex\hbox{E}\kern-.125emX}}
\begin{document}

\newtheorem{innercustomthm}{Theorem}
\newenvironment{customthm}[1]
  {\renewcommand\theinnercustomthm{#1}\innercustomthm}
  {\endinnercustomthm}

\newtheorem{innercustomcor}{Corollary}
\newenvironment{customcor}[1]
  {\renewcommand\theinnercustomcor{#1}\innercustomcor}
  {\endinnercustomthm}
  
\newtheorem{innercustomlemma}{Lemma}
\newenvironment{customlemma}[1]
  {\renewcommand\theinnercustomlemma{#1}\innercustomlemma}
  {\endinnercustomlemma}

\newcommand{\ming}[1]{{\color{blue} #1}}

\newtheorem{theorem}{Theorem}
\newtheorem{fact}{Fact}
\newtheorem{lemma}[theorem]{Lemma}
\newtheorem{corollary}[theorem]{Corollary}
\newtheorem{claim}{Claim}
\newtheorem{proposition}[theorem]{Proposition}
\newtheorem{observation}{Observation}
\newtheorem{definition}{Definition}

\title{Privacy-preserving Blockchain-enabled Parametric Insurance via Remote Sensing and IoT}

\author{
    \IEEEauthorblockN{Mingyu Hao\IEEEauthorrefmark{1}, Keyang Qian\IEEEauthorrefmark{2}, Sid Chi-Kin Chau\IEEEauthorrefmark{3}\thanks{Email: mingyu.hao@anu.edu.au, keyang.qian@monash.edu, Corresponding Author: sid.chau@acm.org}}
    \IEEEauthorblockA{\IEEEauthorrefmark{1}School of Computing, Australian National University}

    \IEEEauthorblockA{\IEEEauthorrefmark{2}Faculty of Information Technology, Monash University}
    \IEEEauthorblockA{\IEEEauthorrefmark{3}Data61, CSIRO}
}

\maketitle

\begin{abstract}
Traditional Insurance, a popular approach of financial risk management, has suffered from the issues of high operational costs, opaqueness, inefficiency and a lack of trust. Recently, blockchain-enabled {\em parametric insurance} through authorized data sources (e.g., remote sensing and IoT) aims to overcome these issues by automating the underwriting and claim processes of insurance policies on a blockchain. However, the openness of blockchain platforms raises a concern of user privacy, as the private user data in insurance claims on a blockchain may be exposed to outsiders. In this paper, we propose a privacy-preserving parametric insurance framework based on succinct zero-knowledge proofs (zk-SNARKs), whereby an insuree submits a zero-knowledge proof (without revealing any private data) for the validity of an insurance claim and the authenticity of its data sources to a blockchain for transparent verification. Moreover, we extend the recent zk-SNARKs to support robust privacy protection for multiple heterogeneous data sources and improve its efficiency to cut the incurred gas cost by 80\%. As a proof-of-concept, we implemented a working prototype of bushfire parametric insurance on real-world blockchain platform Ethereum, and present extensive empirical evaluations. 
\end{abstract}
\

\begin{IEEEkeywords}
Blockchain, Remote Sensing, IoT, Privacy, Zero-Knowledge Proofs
\end{IEEEkeywords}

\vspace{-5pt}
\section{Introduction} \label{sec:intro}

Traditional insurance, known as {\em indemnity insurance}, relies on case-by-case assessments to determine financial losses. The assessment processes typically involve significant manual administration for paperwork validation and approval. Hence, traditional insurance suffers from the issues of (1) high operational costs (because of the manual administration), (2) opaqueness and biases (for the case-by-case assessments), (3) a lack of trust from clients, and (4) inefficiency and delays. These issues are causing rising insurance fees and declining customer satisfaction.

\subsection{Parametric Insurance} 

To address the issues of traditional insurance, there is an emerging trend of automating the insurance processes by a data-driven approach. Unlike traditional insurance relying on case-by-case assessments, a new approach called {\em parametric insurance} \cite{lin2020application} determines the validity of a claim and its insurance payout based on a verifiable index, which is usually computed by a publicly known algorithm with publicly available input data. In particular, the growing availability of sensor data from diverse IoT devices and remote sensing sources has made it possible to provide practical data sources for parametric insurance. For example, the catastrophic insurance offered by Mexican Coastal Management Trust Fund \cite{insuresilience} reimburses losses to the local environment and tourism industry caused by cyclones, based on locally measured wind speed. There are a wide range of parametric insurance products, such as flight delay, bushfire, flooding, crop and solar energy insurance \cite{mingyu2023blochchain}, that can be determined by publicly available sensor data.

Parametric insurance offers several advantages over traditional insurance. First, parametric insurance does not require the demonstration of causation, and the measurement of losses based on publicly available data is more objective. Hence, it simplifies insurance claim processes. Second, parametric insurance uses a publicly known algorithm with publicly verifiable data to improve the transparency and reduce the human biases in case-by-case assessments. Third, parametric insurance streamlines the management of insurers.

\subsection{Blockchain-enabled Parametric Insurance}

More importantly, parametric insurance presents an opportunity of automation through blockchain and authorized data sources (e.g., remote sensing and IoT). In particular, {\em permissionless} blockchain platforms (e.g., Ethereum) ensure transparency/traceability/accountability via an open ledger and smart contracts. The open ledger stores immutable records of transactions and contracts without a centralized manager.

For blockchain-enabled parametric insurance, the insurer initially encodes the insurance policy and the algorithm for calculating the parametric insurance index in a smart contract. The insuree can then submit an insurance claim with supporting data to the smart contract. Subsequently, the smart contract automatically verifies parametric insurance and algorithmically approves or rejects insurance payouts without manual intervention. Consequently, blockchain-enabled insurance minimizes processing time, reduces operational costs, and enhances transparency. There have been several existing parametric insurance providers using blockchain (e.g., Etherisc\cite{Etherisc}).

\subsection{Privacy-preserving Blockchain-enabled Parametric Insurance}

Despite the benefits, there are two major challenges that hinder blockchain-enabled insurance in practice:
\begin{itemize}

\item {\bf Blockchain Privacy}: The transparency of smart contracts mandates that transaction records and processing data be publicly visible and traceable. Even though one may use pseudo-anonymity to hide the identities in the open ledger, it is possible to deanonymize transactions and infer the true identities \cite{biryukov2019deanonymization}. In parametric insurance, every insuree can access the smart contracts and associated data. As a result, one user's claim may be also visible to other users and this compromises user privacy.

\item {\bf Blockchain Processing Cost}: Another significant challenge of using smart contracts is the high computational cost. The execution cost of smart contracts is usually metered by the incurred computation and memory space. For example, on Ethereum platform, each smart contract deployment or execution costs a ``gas fee" \cite{gasandfees}, which can be expensive depending on the memory space and the complexity of its execution. Thus, any computational intensive tasks (e.g., data processing over large satellite imagery) should not be executed by smart contracts.

\end{itemize}
To tackle these challenges, we utilize succinct zero-knowledge proofs to enhance the privacy and efficiency of blockchain computation. A {\em zero-knowledge proof} enables a prover, who possesses a secret, to convince a verifier of their possession of the secret without revealing the secret itself \cite{rackoff1991non}. In particular, Zero-Knowledge Succinct Non-interactive ARguments of Knowledge ({\em zk-SNARKs}) provide compact proofs for efficient verification without any interactions between the prover and verifier, which are widely used in blockchain applications.

\begin{figure}[t]  
	\centering
	  \includegraphics[width=\linewidth]{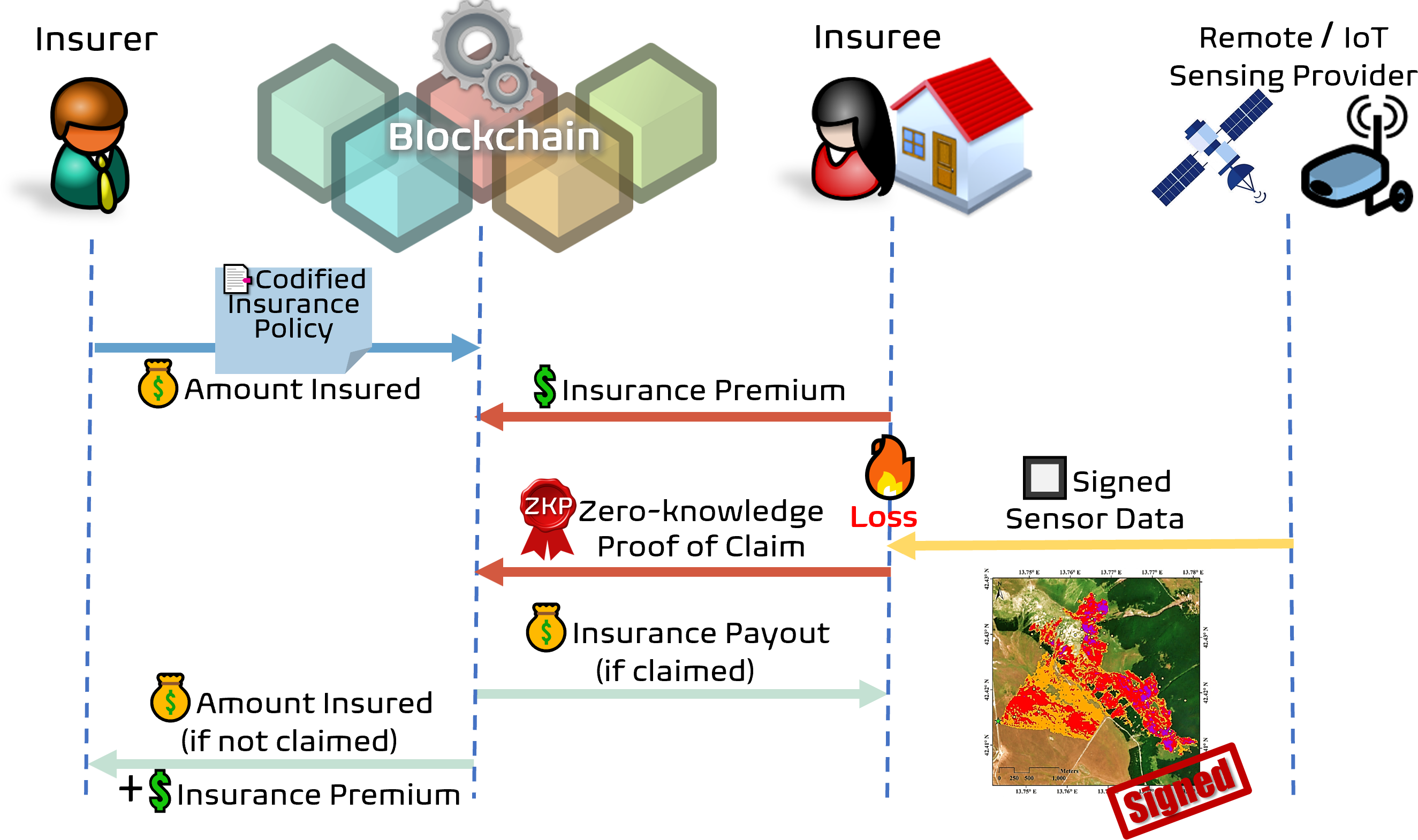} 
	  \caption{An illustration of privacy-preserving blockchain-enabled parametric insurance protocol.}
	  \label{fig:scenario}
\end{figure}

In this paper, we propose a privacy-preserving blockchain-enabled parametric insurance protocol based on zk-SNARKs for protecting insurees' private data, and enabling efficient on-chain claim verification of insurance policies. The basic idea of our protocol is illustrated in Fig.~\ref{fig:scenario}. First, the insurer sets up the amount insured on blockchain by cryptocurrency and the insurance policy by a smart contract. The insuree confirms the policy by paying the insurance premium to the smart contract. To claim the insurance (e.g., for bushfire), the insuree obtains signed sensor data from authorized data providers, and then submits a zero-knowledge proof by zk-SNARK protocol to prove the validity of the claim and the authenticity of its data sources. After successful verification by smart contract on blockchain, the insuree will receive the insurance payout.

In this work, we focus on the application of parametric bushfire insurance, which requires complex processing of satellite images. However, our framework can be generalized to general parametric insurance via remote sensing and IoT.

\noindent
{\bf Contributions}.
Our contributions are summarized as follows:
\begin{enumerate}
    \item We extend a recent zk-SNARK protocol (called Sonic \cite{maller2019sonic}) with the following extensions in Sec.~\ref{sec:extensions}:
    \begin{itemize}
	
        \item \textbf{Heterogeneity and Independence of Data Sources}: It allows a verifier to validate data from heterogeneous authorized sources who may authenticate the data independently and are agnostic to the applications of the data. We propose a zk-SNARK protocol to flexibly incorporate data from heterogeneous IoT or remote sensing providers into blockchain-enable applications. 
		
        \item \textbf{Efficient On-chain Processing}: Regardless of the complexity of the prover's computation, the verifier can verify the proof efficiently with small verification time and proof size. Moreover, we propose an enhanced polynomial commitment scheme to allow batch verification and reduce the incurred verification gas cost on blockchain by up to 80\%. 
		
    \end{itemize} 
	
    \item We propose a novel generic framework for parametric insurance application based on our zk-SNARK protocol in Sec.~\ref{sec:model}. Our framework can be applied to a generic parametric insurance applications. 
	
    \item As a proof-of-concept, we implemented a prototype
    of bushfire insurance on Ethereum blockchain platform, and present extensive empirical evaluations in Sec.~\ref{sec:eval}.	
	
\end{enumerate}

\noindent
{\bf Outline}.
We present the related work in Sec.~\ref{sec:related} and some cryptographic preliminaries in Sec.~\ref{sec:prelim}. Then we explain the ideas of zk-SNARKs and Sonic in Sec.~\ref{sec:zksnarks}. We present our extensions in Sec.~\ref{sec:extensions} and apply our zk-SNARK protocol to the application of parametric bushfire insurance in Sec.~\ref{sec:model}, with evaluations in Sec.~\ref{sec:eval}.

\section{Related Work} \label{sec:related}

\subsection{Blockchain-based Insurance}

Given the promising potential of blockchain technology, the insurance industry has established B3i (Blockchain Insurance Industry Initiative) and blockchain-based insurance platforms with reusable insurance models \cite{b3i}. These platforms often leverage external data accessed via oracle services, such that smart contracts can validate insurance conditions automatically and approve claims without human intervention. For instance, Etherisc provides flight delay insurance \cite{Etherisc} by tracking flight statuses automatically.
But the extant systems rely on {\em weak} privacy protection approaches:
\begin{enumerate}
    \item {\em Pseudo-anonymity}: Insurees generate pseudo-anonymous identifiers whenever they interact with smart contracts on blockchain. However, this approach can be easily deanonymized \cite{biryukov2019deanonymization} by tracing transaction connections through exchanges. Additionally, it does not conceal sensitive blockchain data from public scrutiny.

    \item {\em Permissioned Blockchain}: Insurance applications can be deployed on permissioned blockchain with limited accessibility of sensitive data \cite{he2018toward}, and managed by Attribute-Based Access Control (ABAC). However, limiting accessibility also reduces accountability and transparency. Also, users' private data can still be accessible to others who interact with the same insurance smart contracts.

\end{enumerate}

To the best of our knowledge, there is no other existing decentralized insurance solution that provides strong privacy protection, while allowing transparent on-chain verification of insurance claims.

\subsection{zk-SNARKs}
\label{sec:relatedWork_zkp}

The concept of zk-SNARKs (Zero-Knowledge Succinct Non-interactive Arguments of Knowledge) began with decades of research in interactive proof systems \cite{cormode2012practical,blumberg2014verifiable,Flashproofs} and probabilistic checkable proofs \cite{thaler2022proofs}. Although the theoretical possibility of zk-SNARKs has been shown by applying classical PCP theorems and Merkle trees, practical constructions of zk-SNARKs only began recently with Pinocchio \cite{parno2016pinocchio}. Since then, there have been numerous advances in realizing practical zk-SNARKs \cite{gabizon2019plonk,ben2018fast,boneh21kzg,groth2016size}.

Practical zk-SNARKs can be broadly classified as follows:
\begin{enumerate}

    \item {\em Trusted Setup}: This class of zk-SNARKs requires a trusted party to set up certain public parameters for a prover to construct a proof \cite{thaler2022proofs}. The public parameters may be generated from a trapdoor (i.e., secret information), such that, without knowing the trapdoor, it would be computationally hard for a false statement to pass the verification. Note that the presence of a trusted setup greatly simplifies the verification of zk-SNARKs, yielding compact proofs and efficient verification. 
	
	{\bf\em Remarks:} Trusted setup does {\em not} necessarily entail weaker security. In fact, in certain applications, there is always a natural party for generating the public parameters, who has no incentive to compromise its trapdoor. Particularly, in insurance applications, the insurer is a natural party for generating the public parameters to allow an insuree to prove the validity of an insurance claim, as long as the insurance policy, once agreed by both parties, can only benefit the insuree (but not the insurer) if the insurance claim is proved valid. In this case, even though the insurer knows a trapdoor to prove false insurance claims, he has no incentive to leak it to any insuree\footnote{An insurance policy between an insuree and the insurer should also be independent of the policies of other insurees. The independence can be coded in the smart contract of insurance policy that can be checked by an insuree before accepting the insurance policy.}. 

    There are two classes of zk-SNARKs with trusted setup: 
	\begin{enumerate}

    \item {\em Circuit-specific Setup}: This class requires circuit-specific public parameters. Namely, the public parameters depend on the circuit structure of a statement. Different circuits will require new public parameters. This class of zk-SNARKs yields the smallest proofs and most efficient verification. Pinocchio \cite{parno2016pinocchio} and Groth16 \cite{groth2016size} are examples with circuit-specific setup.

    \item {\em Circuit-universal Setup}: Another more flexible class of zk-SNARKs has universal public parameters for every circuit. Unlike circuit-specific zk-SNARKs such as Groth16 \cite{groth2016size}, the public parameters can be set up independent of a statement, which significantly reduces the computation cost. Sonic \cite{maller2019sonic} is a zk-SNARK protocol that supports a universal and continually updatable structured reference string (SRS) that scales linearly in size. It has a constant proof size and verification time. Some recent protocols \cite{chiesa2020marlin,bunz2020transparent} are based on Sonic but with different approaches to validating the circuit computation. Plonk \cite{gabizon2019plonk}, on the other hand, uses fan-in-two gates with unlimited fan-out circuits to encode the problem, leading to a more flexible circuit structure and smaller universal setup overhead. It also further reduced the verification cost. 
\end{enumerate}
    \item {\em Transparent Setup}: This class of zk-SNARKs does not require a trusted party to set up the public parameters. The public parameters can be generated without a trapdoor. Hence, the setup can be established transparently \cite{ben2018scalable,lee2021dory,thaler2022proofs}. zk-SNARKs with a transparent setup (so-called zk-STARKs) is useful for certain applications, such that there is no natural party who does not have a conflict of interest in proving a false statement (e.g., decentralized finance). It is not straightforward to decompose the input from the logic in the circuit construction. We note that zk-SNARKs with transparent setup typically require larger proofs and higher verification computation, and hence, are less practical for deployment on real-world permissionless blockchain platforms. 
	
\end{enumerate}

{\bf\em Remarks:} As we will show in later sections, our insurance application requires certain independence between the input sources (e.g., satellite image providers) and circuit designer (i.e., insurer who designs the insurance policy). Namely, the satellite image providers should generate authorized satellite images without knowing how the data is used by third parties. Otherwise, there may lead to possible collusion between the satellite image providers and the insurer, which will undermine the integrity of blockchain-enabled parametric insurance. Sonic inherently enables independence between the circuits and input sources and does not require hard-coding of the labelling of inputs in zk-SNARK construction. This is important in our application, where the input sources and circuit are required to be defined by separate independent parties.

On the other hand, Plonk-based protocols involve a mapping from the logic gate IDs to the corresponding logic gate values. 
In our application, the input source is required to commit the input data independently from the circuit, without knowing the gate ID mapping. Achieving such a level of independence between the input data and circuit structure is straightforward in Sonic, which may be challenging in Plonk.
Also, it is not straightforward to perform authentication on the input sources separately from the circuit. Particularly, this may require coordination between the input sources and circuit designer on the mapping from the gate IDs to the corresponding gate values, which should be defined independently by the satellite image providers and circuit designer to preclude possible collusion. Therefore, there is an advantage of Sonic over Plonk for allowing validated input data from independent data sources. Hence, we adopt Sonic as the basic framework in this paper.

\section{Cryptographic Preliminaries} \label{sec:prelim}

In the following, we briefly present the basic cryptographic preliminaries used in zk-SNARKs, before explaining the basic ideas of zk-SNARKs and our protocol in the subsequent sections. More detailed cryptographic preliminaries can be found in standard cryptography textbooks \cite{crytobk}. 

First, ${\mathbb F}_p = \{0, ..., p-1\}$ denote a finite field of integers modulo $p$. We write ``$x+y$'' and ``$x y$'' for modular arithmetic without explicitly mentioning ``${\tt mod\ } p$''. We consider a cyclic group ${\mathbb G}$ of prime order $p$ (e.g., an elliptic curve group). Let ${\tt g}$ be a generator of ${\mathbb G}$, such that ${\tt g}$ can generate any element in ${\mathbb G}$ by taking proper powers (i.e., for each ${\tt k} \in {\mathbb G}$, there exists $x \in {\mathbb F}_p$ such that ${\tt k} = {\tt g}^x$). We write $x \xleftarrow{\$}\mathbb {\mathbb F}_p$ to mean selecting $x$ in ${\mathbb F}_p$ at uniformly random. The {\em computational Diffie-Hellman assumption} states that given ${\tt g}^x$, it is computationally hard to obtain $x$, which underlies the security of many crypto systems.

\subsection{Bilinear Pairing}

A useful property of elliptic curve groups is bilinear pairing. A {\em bilinear pairing} is a mapping ${\tt e}:\mathbb G_1 \times \mathbb G_2 \mapsto \mathbb G_T$, where $\mathbb G_1, \mathbb G_2$ are cyclic groups of prime order $p$, such that
$${\tt e}\langle {\tt g}^x, {\tt h}^y\rangle = {\tt e}\langle {\tt g}, {\tt h}\rangle^{xy}$$
for any ${\tt g} \in \mathbb G_1, {\tt h} \in \mathbb G_2$, and $x,y \in \mathbb F_p$. There are pairing-friendly groups that admit efficient bilinear pairing \cite{crytobk}.	

A key consequence of pairing is that given $({\tt g}^x, {\tt h}^y, {\tt k})$, a verifier can verify whether ${\tt k} \overset{?}{=} {\tt g}^{xy}$ by checking the following:
\begin{equation}\label{eqn:pairing}
{\tt e}\langle {\tt k}, {\tt h} \rangle \overset{?}{=} 
{\tt e}\langle {\tt g}^x, {\tt h}^y\rangle  ={\tt e}\langle {\tt g}, {\tt h}\rangle^{xy} ={\tt e}\langle {\tt g}^{xy}, {\tt h} \rangle 
\end{equation}
By the computational Diffie-Hellman assumption, the above verification does not need to reveal $(x, y)$, which may be used to represent some private data.

\subsection{Polynomial Commitment}

A polynomial\footnote{In this paper, we denote indeterminate variables by capital letters $X, Y$.} $f[X]$ can represent enormous information. For example, one can represent a sequence $(a_i)_{i=0}^{d}$ using a polynomial $f[X]$, by expressing $(a_i)_{i=0}^{d}$ as $f$'s coefficients: $f[X] = \sum_{i=0}^{d} a_i X^i$. In the next section, we will represent a decision problem by a polynomial.

A (univariate) polynomial commitment scheme allows a prover to commit to a univariate polynomial (as a secret) in advance and to open the evaluations at specific values subsequently with a proof to show that the evaluated polynomial is identical in the commitment. A polynomial commitment provides confidence that the prover does not cheat. A generic polynomial commitment scheme consists of four methods (\textsf{\small Setup}, \textsf{\small Commit}, \textsf{\small Open}, \textsf{\small Verify}) which are explained in Table~\ref{tab:polycm}.

There are several desirable properties of a polynomial commitment scheme:

\begin{itemize}
    \item \textbf{Correctness}: If $F \leftarrow \textsf{\small Commit}({\tt srs}, f[X])$ and $(v, \pi) \leftarrow \textsf{\small Open}({\tt srs}, F, z)$, then $\textsf{\small Verify}({\tt srs}, F, z, v, \pi) = {\tt True}$.

    \item \textbf{Knowledge Soundness}: 
    For every successful polynomial time adversary $\mathcal A$, there exists an efficient extractor ${\mathcal E}_{\mathcal A}$ who can extract the polynomial with high probability given the access to the adversary ${\mathcal A}$'s internal states: 
	\begin{align*}
	{\mathbb P}&\left [ \begin{array}{l}
             \textsf{\small Verify}({\tt srs}, F, z, v, \pi) = {\tt True}  \\ \wedge f[z]=v 
        \end{array}     
        \middle|
        \begin{array}{l}
            {\tt srs} \leftarrow \textsf{\footnotesize Setup}({\lambda})\wedge \\ 
            f[X] \leftarrow {\mathcal E}_{\mathcal A}({\tt srs}, \pi)
            \end{array}
        \right]  \\ &= 1 - \epsilon(\lambda) 
	\end{align*}
	where $\epsilon(\lambda)$ is a decreasing function in $\lambda$, such that  $\epsilon(\lambda) \to 0$ (i.e., $\epsilon(\lambda)$ is negligible), when $\lambda \to \infty$.
    \item \textbf{Computational Hiding}: No adversary can determine $f[z]$ from commitment $F$ before the evaluation at $z$ is revealed, with high probability ($1-\epsilon(\lambda)$).
\end{itemize}

\begin{table}[t]
\caption{Generic Polynomial Commitment Scheme} \label{tab:polycm} \vspace{-5pt}
\begin{longfbox}[border-break-style=none,border-color=\#bbbbbb,background-color=\#eeeeee,width=\linewidth]
\begin{itemize}[leftmargin=*]

    \item \textsf{\footnotesize Setup}$({\lambda}) \rightarrow {\tt srs}$: \textsf{\footnotesize Setup} takes the security parameter ${\lambda}$ and outputs a {\em structured reference string} ${\tt srs}$, which is a public parameter to the commitment scheme.\\

    \item \textsf{\footnotesize Commit}$({\tt srs}, f[X]) \rightarrow F$: \textsf{\footnotesize Commit} generates a commitment $F$ given a polynomial $f[X]$. \\

     \item \textsf{\footnotesize Open}$({\tt srs}, F, z) \rightarrow (v, \pi)$: \textsf{\footnotesize Open} evaluates polynomial $f[X]$ with its commitment $F$ at value $X=z$, and outputs evaluation $v=f[z]$ (i.e., an opening), together with a proof $\pi$ to prove the following relation: 
    \begin{align*}
    \mathcal {\mathscr R}_{\rm cm} \triangleq \Big\{ (F, z, v) \mid 
        \textsf{\footnotesize Commit}({\tt srs}, f[X])= F \wedge f[z]=v \Big\}  
    \end{align*}

    \item \textsf{\footnotesize Verify}$({\tt srs}, F, z, v, \pi) \rightarrow \{{\tt True},{\tt False}\}$: \textsf{\footnotesize Verify} uses proof $\pi$ to verify whether $(F, z, v)  \in {\mathscr R}_{\rm cm} $. It outputs ${\tt True}$, if the verification is passed, otherwise ${\tt False}$.
		
\end{itemize}
   
\end{longfbox} \vspace{-5pt}
\end{table}

\subsection{KZG Polynomial Commitment}\label{sec:kzg}

A concrete realization of a polynomial commitment scheme is KZG polynomial commitment scheme \cite{kate2010constant}, which is being incorporated in the Ethereum standard EIP-4844\cite{eip4844}.

We generally consider a Laurent polynomial with negative power terms, such that $f[X] = \sum_{i = - d}^{d} a_i X^i$, and naturally extend polynomial commitment schemes to Laurent polynomials. Lemma~\ref{lemma:factor} is a basic fact about factoring a polynomial.

\begin{lemma}\label{lemma:factor}
   Given a Laurent polynomial $f[X]$ and value $z$, then the polynomial $f[X]-f[z]$ is divisible by $X-z$, namely,  $f[X]-f[z] = (X-z) \cdot q[X]$ for some Laurent polynomial $q[X]$. Intuitively, it is because $X=z$ is a root of $f[X]-f[z]$.
\end{lemma}

KZG polynomial commitment scheme is a concrete polynomial commitment scheme that can be verified efficiently with constant time complexity. Based on Eqn.~(\ref{eqn:pairing}) and Lemma~\ref{lemma:factor}, we specify the four methods of KZG polynomial commitment scheme ($\textsf{\small Setup}_{\tt KZG}, \textsf{\small Commit}_{\tt KZG}, \textsf{\small Open}_{\tt KZG}, \textsf{\small Verify}_{\tt KZG}$) in Table~\ref{tab:kzg}. KZG polynomial commitment scheme is shown to satisfy correctness, knowledge soundness, and computational hiding under the computational Diffie-Hellman assumption \cite{kate2010constant}.

{\bf\em Remarks:} Maller et al. proposed a stronger version of KZG polynomial commitment scheme to adapt KZG commitment into Sonic protocol \cite{maller2019sonic}. Their scheme satisfies {\em bounded polynomial extractability}, such that there is an extractor to extract $f[X]$ of degree $d'$ from the proof $\pi$, if $\deg(f[X]) = d' < d$ is known in advance.

KZG polynomial commitment scheme requires a trusted party for the setup of structured reference string ${\tt srs}$. However, this is not an issue in our application of parametric insurance, because the insurer can set up ${\tt srs}$ to let the clients prove the validity of their claims, and the insurer has no incentive to compromise ${\tt srs}$, if this does not benefit the insurer.

\begin{table}[t] 
\caption{KZG Polynomial Commitment Scheme (${\tt KZG}$)} \label{tab:kzg} \vspace{-5pt}
\begin{longfbox}[border-break-style=none,border-color=\#bbbbbb,background-color=\#eeeeee,width=\linewidth] 
\begin{itemize}[leftmargin=*]

    \item $\textsf{\footnotesize Setup}_{\tt KZG}$: We suppose that there is a trusted party, who takes the security parameter ${\lambda}$ and generates $\mathbb G_1, \mathbb G_2, \mathbb G_T$ with bilinear pairing ${\tt e}$. Then, it selects ${\tt g} \in {\mathbb G_1}, {\tt h} \in {\mathbb G_2}, {x} \xleftarrow{\$} {\mathbb F_p}\backslash\{0, 1\}$ at random uniformly. Next, set the structured reference string as: 
	$${\tt srs} \leftarrow \big({\tt e}, ({\tt g}^{{x}^i})_{i = - d}^{d}, {\tt h}, {\tt h}^{x} \big)$$

    \item $\textsf{\footnotesize Commit}_{\tt KZG}$:  Given a Laurent polynomial $f[X] = \sum_{i = - d}^{d} a_i X^i$ with non-zero constant term, set the commitment as: 
	$$F \leftarrow {\tt g}^{f[x]} = \prod_{i=-d}^{d}  ({\tt g}^{{x}^i})^{a_i}$$

    \item $\textsf{\footnotesize Open}_{\tt KZG}$: To generate a proof $\pi$ to the evaluation $v=f[z]$ for commitment $F$, compute polynomial $q[X] = \frac{f[X]-f[z]}{X-z}$ by a polynomial factorization algorithm. Suppose $q[X] = \sum_{i = - d}^{d} b_i X^i$. Then, set the proof by: 
	$$\pi  \leftarrow  {\tt g}^{q[x]} = \prod^{d}_{i=-d}  ({\tt g}^{{x}^i})^{b_i}$$
	    
    \item $\textsf{\footnotesize Verify}_{\tt KZG}$:  To verify $({\tt srs}, F, z, v, \pi)$, the verifier checks the following pairing equation:
	$${\tt e}\langle F\cdot {\tt g}^{-v}, {\tt h}\rangle \overset{?}{=} {\tt e}\langle \pi, {\tt h}^{x} \cdot {\tt h}^{-z}\rangle$$
	That is, checking ${\tt e}\langle {\tt g}, {\tt h}\rangle^{f[x] - v} \overset{?}{=} {\tt e}\langle {\tt g}, {\tt h}\rangle^{(x-z)\cdot q[x]}$, which follows from Eqn.~(\ref{eqn:pairing}) and Lemma~\ref{lemma:factor}.

\end{itemize}
\end{longfbox} \vspace{-5pt}
\end{table}

\subsection{Digital Signature} \label{sec:background_DSA}

A digital signature scheme can be used to authenticate a data source, which consists of three methods (\textsf{\small Setup}, \textsf{\small Sign}, \textsf{\small VerifySign}), as explained in Table~\ref{tab:sign}. Assume that the public key is shared through a secure channel. There are several key properties of a digital signature scheme:
\begin{itemize}
    \item \textbf {Authenticity}: A signature generated by the secret key and deliberately signed on some message will always be accepted using the corresponding public key. 

    \item \textbf{Unforgeability}: Given a message and a public key, it is impossible to forge a valid signature on the message without knowing the secret key. 

    \item \textbf{Non-reusability}: It is infeasible for a single signature to pass the verifications of two different messages. 
\end{itemize}
Common RSA encryption or Elliptic Curve Digital Signature (ECDSA) based digital signatures can satisfy authenticity, unforgeability, and non-reusability.

\begin{table}[t]
\caption{Generic Digital Signature Scheme} \label{tab:sign} \vspace{-5pt}
\begin{longfbox}[border-break-style=none,border-color=\#bbbbbb,background-color=\#eeeeee,width=\linewidth]
\begin{itemize}[leftmargin=*]
	\item \textsf{\footnotesize  Setup}$({\lambda})  \rightarrow ({\tt pk, sk})$:  \textsf{\footnotesize  Setup} takes a security parameter ${\lambda}$ and outputs an asymmetric key pair $({\tt pk, sk})$. \\ 

	\item \textsf{\footnotesize Sign}$({\tt sk}, m)  \rightarrow \sigma$: \textsf{\footnotesize Sign} takes a message $m$ with a fixed length, and outputs a signature $\sigma$ using the secret key ${\tt sk}$. \\

	\item \textsf{\footnotesize VerifySign}$({\tt pk}, m, \sigma)  \rightarrow \{{\tt True},{\tt False}\}$:  \textsf{\footnotesize VerifySign} uses the public key ${\tt pk}$ to verify message $m$ and signature $\sigma$. It outputs ${\tt True}$, if the verification is passed, otherwise ${\tt False}$.
\end{itemize}
\end{longfbox}
\end{table}

\section{zk-SNARKs and Sonic Protocol} \label{sec:zksnarks}

Based on the preliminaries in the previous section, this section presents the concept of zk-SNARKs and Sonic protocol, which will be used to construct zero-knowledge proofs for parametric insurance claims in the next section.

\subsection{Arguments of Knowledge and zk-SNARKs}

\begin{table}[t]
\caption{Generic Argument System} \label{tab:arg} \vspace{-5pt}
\begin{longfbox}[border-break-style=none,border-color=\#bbbbbb,background-color=\#eeeeee,width=\linewidth]
\begin{itemize}[leftmargin=*]

    \item \textsf{\footnotesize Setup}$(\lambda, C)  \rightarrow {\tt srs}$: \textsf{\footnotesize Setup} takes a relation ${\mathscr R}_C$ as well as a security parameter $\lambda$ as input, and outputs a {\em structured reference string} containing the public parameters for proof generation and verification. \\

    \item \textsf{\footnotesize Prove}$({\tt srs},x,w) \rightarrow \pi$: \textsf{\footnotesize Prove} uses the common reference string ${\tt srs}$, public input $x$, and the secret witness $w$ to generate a proof $\pi$, which is a proof of the following statement: ``given $C$ and $x$, there exists a secret witness $w$, such that $(x,w)\in {\mathscr R}_C$".  \\

    \item \textsf{\footnotesize Verify}$({\tt srs},x,\pi)  \rightarrow \{{\tt True},{\tt False}\}$: \textsf{\footnotesize Verify} verifies the proof $\pi$, and outputs ${\tt True}$ if $(x, w) \in {\mathscr R}_C$, otherwise ${\tt False}$.
\end{itemize}
\end{longfbox} \vspace{-5pt}
\end{table}

zk-SNARKs belong to a general concept called {\em argument system}, which is a protocol between a prover and a verifier for proving the satisfiability of a statement in a given NP language \cite{snarkintro}. In this paper, we focus on the NP language of satisfiability problems, which is sufficient to encode an insurance policy in the subsequent section. Given a finite field $\mathbb {\mathbb F}_p$, a decision function is denoted by $C :\mathbb {\mathbb F}_p^n \times \mathbb {\mathbb F}_p^m \rightarrow \mathbb {\mathbb F}_p^l$, which takes two parts of input: a public component as public input $x \in \mathbb {\mathbb F}_p^n$ and a secret component as witness $w \in \mathbb {\mathbb F}_p^m$. Define the relation ${\mathscr R}_C \triangleq \{(x, w) \in \mathbb {\mathbb F}_p^n \times \mathbb {\mathbb F}_p^m : C(x, w) = 0\}$. An argument system allows a prover to convince a verifier the knowledge of $(x, w) \in {\mathscr R}_C$, without revealing $w$. 

A generic argument system consists of three methods  (\textsf{\small Setup}, \textsf{\small Prove}, \textsf{\small Verify}), as explained in Table~\ref{tab:arg}. An argument system has several key properties, typically, completeness, knowledge soundness and perfect honest-verifier zero knowledge. {\em Zero-Knowledge Succinct Non-Interactive Arguments of Knowledge} (zk-SNARKs) are argument systems with the additional properties of succinctness and non-interactiveness.

\subsection{Building Blocks of zk-SNARKs}

In this subsection, we introduce the building blocks of a zk-SNARK protocol. In general, zk-SNARK protocols have a general framework with the following components:
\begin{enumerate}

    \item \textbf{Problem Characterization}: The first step is to express a decision problem (e.g., deciding parametric insurance) by a suitable decision function $C$, such that the public input $x$ represents a problem instance (e.g., insurance policy) and the witness $w$ represents the private data (e.g., insurance claim). If $C(x,w)=0$ is satisfiable by $w$, then the decision problem returns true (e.g., the insurance claim is valid). In Sec.~\ref{sec:model}, we will present a concrete example of expressing the problem of bushfire parametric insurance as a decision function.
	
    \item \textbf{Polynomial IOP}: The next step is to map the decision function $C$ to a suitable polynomial $f[X]$, such that the satisfiability of $C$ can be validated by checking certain properties of $f[X]$. For example, if $C(x, w)=0$ is satisfiable by a witness $w$, then $f[X]$ has a zero constant term. However, the degree of $f[X]$ may be very large. To prove the satisfiability of $C$ efficiently, the verifier uses a polynomial interactive oracle proof ({\em polynomial IOP}), such that the verifier only queries a small set of evaluations of $f[X]$ from the prover, rather than obtaining the entire list of coefficients of $f[X]$ from the prover. 

    \item \textbf{Polynomial Commitment}: Note that the prover may be dishonest and the verifier cannot trust the prover for honest evaluations of $f[X]$. Hence, the prover is required to use a polynomial commitment scheme. The prover first needs to send the commitment $F=\textsf{\small Commit}(f[X])$ to the verifier, and then provides an opening and its correct proof of any requested evaluation $(v, \pi)=\textsf{\small Open}(F,z)$.
	
\end{enumerate}

\subsection{Sonic zk-SNARK Protocol}

\begin{figure}[h]  
	\centering
	  \includegraphics[width=\linewidth]{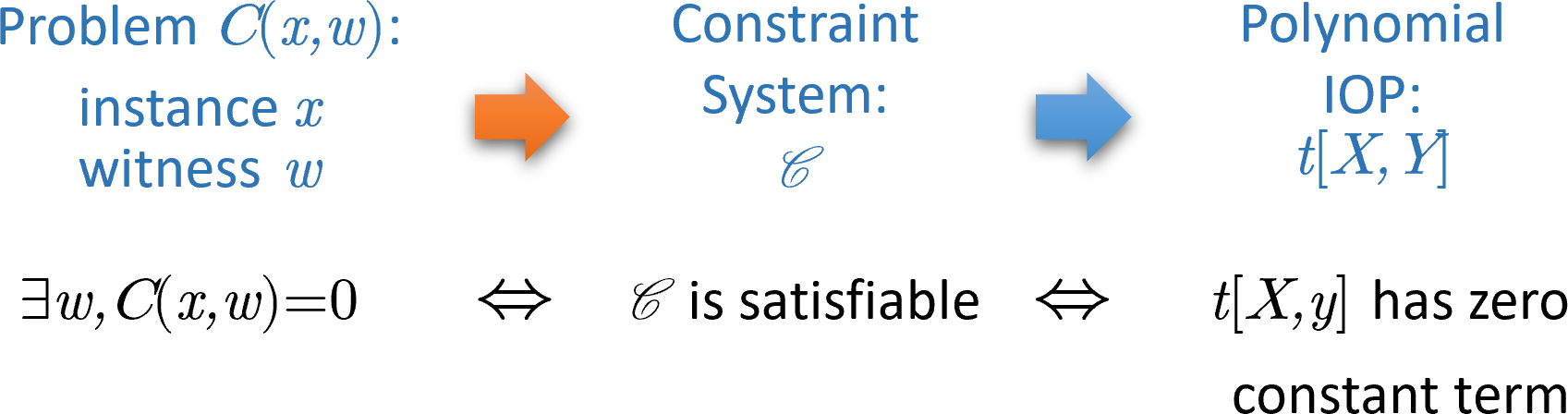} 
	  \caption{The basic construction of Sonic zk-SNARK protocol.}
	  \label{fig:flow}
\end{figure}

In this section, we describe the Sonic zk-SNARK protocol \cite{maller2019sonic}. The basic construction of Sonic is illustrated in Fig.~\ref{fig:flow}. First, we map the decision function $C$ to a constraint system ${\mathscr C}$ that consists of additive and multiplicative constraints. Then, we represent the constraint system ${\mathscr C}$ by a bivariate polynomial $t[X,Y]$. We will show that the satisfiability of $C(x, w) = 0$ is equivalent to the property that the univariate polynomial $t[X,y]$ has a zero constant term at any given $Y=y$.

Note that here we present a simplified Sonic protocol for clarity, with slightly weaker security. But we will explain the original Sonic protocol in the discussion section. We next explain the detailed construction of Sonic protocol as follows.

\subsubsection{Mapping Decision Function to Constraint System}

We denote the $i$-th entry of the vector $\mathbf{a}$ by $\mathbf a_i$. We represent the decision function $C$ by a specific {\em constraint system} (denoted by ${\mathscr C}$) with $N$ multiplicative constraints and $Q$ linear constraints, such that each multiplicative constraint captures a (2-fan-in) multiplication in $C$, whereas each linear constraint captures a (multi-fan-in) addition in $C$ in the following manner:
\[
\left\{
\begin{array}{r@{\ }ll}
\mathbf a_i \cdot \mathbf b_i & = \mathbf c_i, & \mbox{for\ }  1 \le i \le N \\
\mathbf a \cdot \mathbf u_q + \mathbf b \cdot \mathbf v_q + \mathbf c \cdot \mathbf w_q & = k_q, & \mbox{for\ }  1 \le q \le Q \\
\end{array}
\right. \qquad (\mathscr C)
\]
where vectors $\mathbf{a, b, c} \in {\mathbb F}_p^N$ denote the left inputs, the right inputs and the outputs of multiplications in $C$, respectively. 

Note that ${\mathbf u_q, \mathbf  v_q, \mathbf  w_q} \in \mathbb F^N_p$ and $k_q \in {\mathbb F}_p$ capture the specification of a given instance of decision function $C$ (i.e., the public input). The satisfiability of decision function $C$ is equivalent to deciding whether there exist $(\mathbf{a, b, c})$ (i.e., the witness) to satisfy constraint system ${\mathscr C}$, given $({\mathbf u_q, \mathbf  v_q, \mathbf  w_q}, k_q)_{q=1}^Q$.

\subsubsection{Mapping Constraint System to Polynomial IOP} 
\

We next introduce an indeterminate variable $Y\in\mathbb {\mathbb F}_p$. We associate each constraint in constraint system ${\mathscr C}$ with a coefficient in each power term of $Y$ as follows: (1) the $i$-th multiplicative constraint is associated with power term $(Y^i+Y^{-i})$, and (2) the $q$-th linear constraint is associated with power term $Y^{q+N}$. The constraint system ${\mathscr C}$ can be represented by a polynomial ${\mathscr C}[Y]$, as defined as follows:
\begin{align} 
{\mathscr C}[Y]\triangleq& \sum^N_{i=1}(\mathbf a_i \cdot \mathbf b_i - \mathbf c_i)(Y^i+Y^{-i}) \notag \\
& + \sum^Q_{q=1}(\mathbf a\cdot\mathbf u_q+\mathbf b\cdot\mathbf v_q+\mathbf c\cdot\mathbf w_q-k_q)Y^{q+N} 
\end{align}
Polynomial ${\mathscr C}[Y]$ can be further simplified as follows:
\begin{align} 
{\mathscr C}[Y] = & \sum^N_{i=1}(\mathbf a_i \cdot \mathbf b_i - \mathbf c_i)(Y^i+Y^{-i}) \notag \\
& + \Big(\mathbf a\cdot \hat{\mathbf u}[Y]+\mathbf b\cdot \hat{\mathbf v}[Y]+\mathbf c\cdot \hat{\mathbf w}[Y]-\hat{k}[Y]\Big) 
\end{align}
where vectors $\hat{\mathbf u}[Y], \hat{\mathbf v}[Y], \hat{\mathbf w}[Y], \hat{k}[Y]$ are defined as:
\begin{align}
\hat{\mathbf u}_i[Y]\triangleq\ & \sum^Q_{q=1}\mathbf u_{q,i} Y^{q+N},\qquad \qquad \qquad  
\hat{\mathbf v}_i[Y]\triangleq \sum^Q_{q=1}\mathbf v_{q,i} Y^{q+N} \notag
\\
\hat{\mathbf w}_i[Y]\triangleq\ & -Y^i - Y^{-i} + \sum^Q_{q=1}\mathbf w_{q,i} Y^{q+N},\ 
\hat{k}[Y]\triangleq \sum^Q_{q=1}k_{q} Y^{q+N} \notag
\end{align}
Note that ${\mathscr C}$ is satisfiable by $(\mathbf{a, b, c})$, if and only if ${\mathscr C}[y] = 0$ for any $y$. Hence, one can use a random challenge $y$ for ${\mathscr C}[y]$ to test the satisfiability of ${\mathscr C}$. By Schwartz-Zippel Lemma \cite{crytobk}, the probability of a false positivity in such a random test is small $O(\frac{1}{|{\mathbb F}_p|})$, where $|{\mathbb F}_p|$ is the size of the finite field.

Next, instead of representing ${\mathscr C}[Y]$ by a polynomial IOP, we decompose ${\mathscr C}[Y]$ into multiple bivariate polynomials with indeterminate variables $X, Y$, such that the witness $(\mathbf{a, b, c})$ and the public input $({\mathbf u_q, \mathbf  v_q, \mathbf  w_q}, k_q)$ are captured by separate polynomials. This allows us to segregate the private input and the public input in separate polynomial commitments, and verify them separately for heterogeneous data sources.

As in \cite{bootle2016efficient}, we define the following polynomials:
\begin{align}
r[X,Y]\triangleq\ & \sum^N_{i=1} {\mathbf a}_i (XY)^i+\sum^N_{i=1} {\mathbf b}_i (XY)^{-i} +\sum^N_{i=1} {\mathbf c}_i (XY)^{-i-N} \notag\\
s[X,Y]\triangleq\ & \sum^N_{i=1} \hat{\mathbf u}_i[Y]X^{-i}+\sum^N_{i=1} \hat{\mathbf v}_i[Y]X^{i} +\sum^N_{i=1} \hat{\mathbf w}_i[Y]X^{i+N} \notag\\
t[X,Y]\triangleq\ & r[X,1]\big(r[X,Y]+s[X,Y]\big)-\hat{k}[Y]\label{eqn:txy}
\end{align}
where $r[X,Y]$ is the private input known by the prover only, but $s[X,Y],\hat{k}[Y]$ are the public input known by the verifier.

One can check that ${\mathscr C}[Y]$ is exactly the coefficient of the power term $X^0$ in $t[X,Y]$ (i.e., the constant term of $t[X,y]$ equals to ${\mathscr C}[y]$ at any given $Y=y$). Hence, ${\mathscr C}[y]=0$ is equivalent to having a zero constant term in $t[X,y]$.  
\\

\subsubsection{Sonic Protocol} \label{sonic_origin}
\

We next present the following simple protocol to let a prover who knows $r[X,Y]$ to prove $t[X,y]$ has a zero constant term for a given challenge $y$ from a verifier, based on a polynomial commitment scheme:
\begin{enumerate}

\item The prover first commits to $r[X,1]$, and sends its commitment $R$ to the verifier. Note that this is equivalent to committing to $r[X,Y]$, since $r[XY,1] = r[X,Y]$.
\item The verifier sends random challenges $(z,y) \xleftarrow{\$}\mathbb {\mathbb F}_p^2$ to the prover. 

\item The prover then commits to $t[X,y]$, and sends its commitment $T$ to the verifier. The prover also opens $r[z,1]$ and $r[zy,1]$ with polynomial commitment proofs w.r.t. $R$.
\item The verifier computes $s[z,y]$ and $\hat{k}[y]$, and then verifies if $t[z,y]$ has been committed correctly using Eqn.~(\ref{eqn:txy}):
\[
t[z,y] \stackrel{?}{=} r[z,1](r[zy,1]+s[z,y])-\hat{k}[y]
\]
as well as checking the respective polynomial commitment proofs of $r[X,1], t[X,y]$ w.r.t. $R, T$.

\item The verifier checks if $t[X,y]$ has a zero constant term.
\end{enumerate}

The Sonic protocol (${\tt S}_{\tt 0}$) (described in Table~\ref{tab:sonic}) is similar to the above simple protocol, but with several changes:
\begin{enumerate}

\item {\em Additional Blinders}: Because there are three values revealed related to $r[x, y]$, which includes $r[z, 1]$, $r[zy, 1]$, and $\textsf{Commit}({\tt srs},r[X,1])$, we add four blinders to the polynomial with random coefficients and powers ($-2n-1, -2n-2, -2n-3, -2n-4$). Such blinders make the polynomial indistinguishable from any random polynomial given less than four evaluations. 
Since the random blinders are also part of user input (like other coefficients of $r$), it does not affect the rest of the protocol. 
\item {\em Checking Zero Constant Term}: Note that we cannot check if $t[X,y]$ has a zero constant term by checking $t[0,y] \stackrel{?}{=} 0$, because $t[X, y]$ is a Laurent polynomial, which is undefined at $t[0,y]$.
To resolve this issue, we use a restricted version of KZG polynomial commitment scheme (see Appendix A) that precludes a committed polynomial with a non-zero constant term, and hence, forcing the prover to only commit polynomials with zero constant terms. Step (5) in the above protocol can be skipped.

\item {\em Outsourcing to Prover}: The simple protocol requires the verifier to compute $s[z,y]$ and $\hat{k}[y]$. Rather than computing $s[z,y]$ and $\hat{k}[y]$ by the verifier, the computation can be outsourced to the prover or an untrusted helper, since $s[X,Y]$ and $\hat{k}[Y]$ are the public input. To verify outsourced computation, one needs the polynomial commitments of $s_Y[Y] \triangleq s[1,Y]$ and $\hat{k}[Y]$ in the setup (which can be prepared by the insurer in our application). Then, the outsourced helper commits to $s_X[X] \triangleq s[X,y]$, when given a random challenge $y$. The validity of polynomial commitment $s_X[X]$ can be verified by checking the equation: 
\[
s_X[1]\stackrel{?}{=} s[1,y]\stackrel{?}{=} s_Y[y]
\]
The openings of $s[z,y], \hat{k}[y]$ can be verified by checking the respective polynomial commitment proofs of $s_X[X], \hat{k}[Y]$.

\item {\em Converting to Non-interactiveness}:  we describe how to convert the interactive protocol in Table~\ref{tab:sonic} to be non-interactive. We can employ Fiat-Shamir heuristic to replace the verifier-supplied random challenges $(z, y)$ by hash values from the previous commitments: $y\leftarrow{\textsf{\small Hash}}(R), z\leftarrow{\textsf{\small Hash}}(R|T| S_X)$. Assuming the one-wayness of a collision-resistant hash function $\textsf{\small Hash}(\cdot)$, the prover cannot manipulate the commitments and the subsequent random challenges to pass the verification.

\end{enumerate}

{\bf\em Remarks:} The Sonic protocol is shown to satisfy completeness, knowledge soundness, perfect honest-verifier zero knowledge\footnote{The simplified Sonic protocol can be extended to support perfect honest-verifier zero knowledge by incorporating random masking to $r[X,1]$ to make its polynomial commitment have a statistical uniform distribution. See \cite{maller2019sonic}.}, succinctness (with a constant proof size and constant verification time) \cite{maller2019sonic}. There are some differences with the original Sonic protocol. Sonic protocol also considers a slightly different way of outsourcing to an untrusted helper.

\begin{table}[t]
\caption{Simplified Interactive Sonic zk-SNARK Protocol (${\tt S}_{\tt 0}$)} \label{tab:sonic} \vspace{-5pt}
\begin{longfbox}[border-break-style=none,border-color=\#bbbbbb,background-color=\#eeeeee,breakable=true,,width=\linewidth]
\textbf{Public Input}: $\lambda, s[X,Y], \hat{k}[Y]$ \\
\textbf{Prover's Input}: $r[X,Y]$ \\
\textbf{Interactive zk-SNARK Protocol}:
\begin{enumerate}
\item 
{\rm Setup}: ${\tt srs}\leftarrow{\textsf{\scriptsize Setup}}(\lambda)$,
\[
S_Y\leftarrow{\textsf{\scriptsize Commit}}({\tt srs},s[1,Y]),  \ K\leftarrow{\textsf{\scriptsize Commit}}({\tt srs},\hat{k}[Y])
\]

\item 
{\rm Prover} $\Rightarrow$ {\rm Verifier}:
$g_1, g_2, g_3, g_4\xleftarrow{\$}\mathbb {\mathbb F}_p$ 
\[r[X, Y]=r[X,Y]+\sum^4_{1}g_i(XY)^{-i-2N}
\]
\[
R\leftarrow{\textsf{\scriptsize Commit}}({\tt srs},r[X,1])
\]
\item
{\rm Verifier} $\Rightarrow$ {\rm Prover}: $y\xleftarrow{\$}\mathbb {\mathbb F}_p$ {\color{gray}\it// (Fiat-Shamir): $y\leftarrow{\textsf{\scriptsize Hash}}(R)$}

\item
{\rm Prover} $\Rightarrow$ {\rm Verifier}: 
\[
\begin{array}{r@{}l@{}r@{}l}
T&\leftarrow{\textsf{\scriptsize Commit}}({\tt srs},t[X,y]) \\
S_X&\leftarrow{\textsf{\scriptsize Commit}}({\tt srs},s[X,y]) \mbox{\color{gray}\it\  // (Outsourced to Prover)}
\end{array}
\] 
{\scriptsize \ }

\item
{\rm Verifier} $\Rightarrow${\rm Prover}:  $z\xleftarrow{\$}\mathbb {\mathbb F}_p$ {\color{gray}\it// (Fiat-Shamir): $z\leftarrow{\textsf{\scriptsize Hash}}(R|T| S_X)$}

\item
{\rm Prover} $\Rightarrow${\rm Verifier}: 

{\color{gray}\it// Evaluate the following openings and generate their proofs: }\\
{\color{gray}\it// \quad $r_1= r[z,1],r_2= r[zy,1],t= t[z,y],k=\hat{k}[y]$}\\
{\color{gray}\it// \quad $s= s_X[z],s_1= s_X[1],s_2= s_Y[y]$}
\[
\begin{array}{r@{}l@{\ }r@{}l}
(r_1, \pi_{r_1})&\leftarrow{\textsf{\scriptsize Open}}({\tt srs},R,z),&
(r_2, \pi_{r_2})&\leftarrow{\textsf{\scriptsize Open}}({\tt srs},R,zy) \\ 
(t, \pi_t)&\leftarrow{\textsf{\scriptsize Open}}({\tt srs},T,z) 
\end{array}
\]
{\color{gray}\it// (Outsourced to Prover): }
\[
\left\{
\begin{array}{r@{}l@{\ }r@{}l}
(k, \pi_k)&\leftarrow{\textsf{\scriptsize Open}}({\tt srs},K,y), &
(s, \pi_{s})&\leftarrow{\textsf{\scriptsize Open}}({\tt srs},S_X,z) \\ 
(s_1, \pi_{s_1})&\leftarrow{\textsf{\scriptsize Open}}({\tt srs},S_X,1), &
(s_2, \pi_{s_2})&\leftarrow{\textsf{\scriptsize Open}}({\tt srs},S_Y,y) \\  
& t \leftarrow{r_1(r_2+s)-k}
\end{array}
\right.
\]

\item
Verifier checks:

{\color{gray}\it// Verify: $\big(t[z,y] \stackrel{?}{=} r[z,1](r[zy,1]+s[z,y])-\hat{k}[y]\big)\wedge$}\\
{\color{gray}\it// \qquad \quad $(s_X[1]\stackrel{?}{=} s[1,y]\stackrel{?}{=} s_Y[y])$}\\
{\color{gray}\it// and the respective polynomial commitment proofs}
\[
\begin{array}{r@{}l}
 \big(t \stackrel{?}{=} r_1(r_2+s)-k\big) &\wedge (s_1 \stackrel{?}{=} s_2) \wedge  {\textsf{\scriptsize Verify}}({\tt srs},T,z,t,\pi_t) \wedge \\
 {\textsf{\scriptsize Verify}}({\tt srs},R,z,r_1,\pi_{r_1}) & \wedge {\textsf{\scriptsize Verify}}({\tt srs},R,zy,r_2,\pi_{r_2}) \wedge \\
 {\textsf{\scriptsize Verify}}({\tt srs},K,y,k,\pi_k) & \wedge {\textsf{\scriptsize Verify}}({\tt srs},S_X,z,s,\pi_{s}) \wedge \\
 {\textsf{\scriptsize Verify}}({\tt srs},S_X,1,s_1,\pi_{s_1}) & \wedge {\textsf{\scriptsize Verify}}({\tt srs},S_Y,y,s_2,\pi_{s_2})
\end{array}
\]

\end{enumerate}    
\end{longfbox} \vspace{-5pt}
\end{table}

\section{Extensions of Sonic Protocol} \label{sec:extensions}

In this section, we present two novel extensions to the original Sonic zk-SNARK protocol.

\subsection{Authenticating Heterogeneous Data Sources} \label{sec:dataprotocol}

In a real-world application, the authenticity of its data sources is as important as the correctness of its computation. Some parts of the witness may be from different data sources (e.g., different remote sensing or IoT sensing providers). In parametric insurance, an insurance claim is valid, only if using authenticated data. In addition to proving the satisfiability of witness, we also need to validate the authenticity of the data. For example, an insuree needs to prove that there is a bushfire indicated in the satellite image at the correct time and location from an authenticated remote sensing provider. The original Sonic protocol does not consider the validation of data sources.  

To authenticate the data sources, one may ask the data providers to sign their data, and the signatures will be checked by a verifier for authenticity, along with the verification of the computation. Since KZG polynomial commitments are able to encode general data, we assume that data providers commit the data to KZG polynomial commitments, together with proper signatures on the commitments. A prover, after retrieving the data from a data provider, will incorporate the KZG polynomial commitments and the respective signatures in their proofs of the computation on the authenticated data.

However, there are a few caveats about the data providers:
\begin{itemize}

\item {\em Heterogeneity of Data Sources}: Each data provider is unaware of each other. They do not coordinate to use the same structured reference strings. A prover who gathers the data from different data providers as the input to the computation needs to incorporate separate KZG polynomial commitments from different structured reference strings into the final proofs.

\item {\em Independence of Data Sources}: The data providers are agnostic to the applications of their data. For example, the data providers may be public data repositories, who provide data for general applications. The independence of data providers from the data applications is critical to the impartiality of the data sources (particularly for insurance). Therefore, the data providers do not use the same structured reference strings as in the data applications.

\end{itemize}

We provide an extension to Sonic protocol to address these issues. We consider $J$ data sources, and each data source $j \in \{1, ..., J\}$ has a data sequence denoted by ${\mathbf d}_{j} = ({\mathbf d}_{j,t})_{t=1}^{m_j}$, where $m_j$ is the length of the data sequence.  The data source $j$ commits its data sequence to a polynomial $d_j[X] = \sum_{i=0}^{m_j} {\mathbf d}_{j,t} X^t$. Let the respective polynomial commitment be $D_j = {\textsf{\small Commit}}({\tt srs}_j,d_j[X])$, where ${\tt srs}_j$ is $j$'s specific reference strings, and the respective signature is denoted by $\sigma_j = {\textsf{\small Sign}}({\tt sk}_j,D_j)$ from $j$'s key pair $({\tt sk}_j, {\tt pk}_j)$.

Assume that the prover's witness consists of two parts: (1) the input data from $J$ data sources, (2) the intermediate data in the computation. We encode the witness in a specific format. The input data from the data sources is encoded in vector ${\mathbf a}$ only, and we pad zeros in ${\mathbf b}, {\mathbf c}$ to keep the same length:
\[
{\mathbf a} = (\tilde{\mathbf a},{\mathbf d}_1,...,{\mathbf d}_J), {\mathbf b} = (\tilde{\mathbf b},0,...,0), {\mathbf c} = (\tilde{\mathbf c},0,...,0)
\]
We denote the intermediate data in the computation by $(\tilde{\mathbf a}, \tilde{\mathbf b}, \tilde{\mathbf c})$. Also, we denote the polynomial for encoding $(\tilde{\mathbf a}, \tilde{\mathbf b}, \tilde{\mathbf c})$ by $\tilde{r}[X,Y]$, which also includes the random blinders as described in Section \ref{sonic_origin}. However, because there are in total $5+2j$ evaluations of $r[X, Y]$ revealed ($2+j$ commitments and $3+j$ openings), we set $6+2j$ random blinders to preserve the security of the protocol:
\[
\tilde{r}[X,Y] = \tilde{r}[X,Y] + \sum^{6+2j}_{i=1} \tilde{\mathbf g}_i (XY)^{-i-2N}  \notag\\
\]

Hence, $r[X,Y]$ is re-expressed as:
\[
r[X,Y] = \tilde{r}[X,Y] + \sum_{j=1}^J d_j[XY] (XY)^{N + \sum_{j=1}^{J-1} m _j}
\]

We present an enhanced Sonic protocol (${\tt S}_{\tt dat}$) in Table~\ref{tab:datasource} to incorporate verification of heterogeneous data sources $\{d_j[X]\}$. In ${\tt S}_{\tt dat}$, the prover needs to provide commitments of $\{d_j[X]\}$ as well as their signatures to prove soundness of data commitments. Then, the prover opens $\tilde{r}[z,1]$ and $d_j[z]$ for each data source $j$ with a commitment proof. Using these openings that have been proved to be authentic, the verifier can validate if the data sources are incorporated authentically in the Sonic proof calculation by checking the following equation: 
\[
r[z,1] \stackrel{?}{=} \tilde{r}[z,1] + \sum_{j=1}^J d_j[z] z^{N + \sum_{j=1}^{J-1} m _j}
\]
Note that any modification of input data $d_j$  will be invalidated by the opening of $d_j$ in its KZG polynomial commitment and the signature $\sigma_j$ of the commitment. 

The rest of the protocol is similar to the one in Table~\ref{tab:sonic}. We leave the security proof of the protocol in \cite{HQC24full}.

\subsection{Polynomial Commitments with Batch Verification} \label{sec:enhanced-sonic}

Sonic protocol requires the openings of multiple polynomials at different evaluation points at the verification stage. Normally, this is implemented by separate verification operations. However, it is possible to reduce the verification overhead by batch verification of multiple openings together. We propose a new polynomial commitment scheme specifically designed for Sonic to allow simultaneous openings of multiple polynomials at different evaluation points. Formally, we consider a set of $K$ polynomials $\{ f_i[X] \}_{i=1}^K$. The prover commits to these polynomials as $F_i = \textsf{\footnotesize Commit}({\tt srs}, f_i[X])$.  There are a set of evaluation points ${\mathcal S} = \{z_1, ..., z_n \}$. Only a subset of evaluation points ${\mathcal S}_i \subset {\mathcal S}$ are evaluated on the $i$-th polynomial $f_i[X]$. The prover can open $\{ (f_i[z])_{z \in {\mathcal S}_i} \}_{i=1}^K$ in a batch to the verifier with one single proof to prove all the openings are correct with respect to the commitments $(F_i)_{i=1}^K$. 
In Appendix B, we describe a new polynomial commitment scheme with a specific requirement in Sonic to preclude a committed polynomial with a non-zero constant term. In \cite{HQC24full}, we also describe an enhanced Sonic protocol with batch verification (${\tt S}_{\tt EV}$). 

\begin{table}[t]
\caption{Enhanced zk-SNARK Protocol with Heterogeneous Data Sources (${\tt S}_{\tt dat}$)} \label{tab:datasource} \vspace{-5pt}
\begin{longfbox}[border-color=\#bbbbbb,background-color=\#eeeeee]
\textbf{Public Input}: $\lambda, \hat{s}[X,Y], \hat{k}[Y], ({\tt pk}_j)_{j=1}^J$ \\
\textbf{Data Source $j \in \{1, ..., J\}$}: $d_j[X], {\tt sk}_j$ \\
\textbf{Prover's Input}: $r[X,Y]$ \\
\textbf{Interactive zk-SNARK Protocol}:
\begin{enumerate}
\item 
{\rm Setup}: ${\tt srs},\  {\tt srs}_j\leftarrow{\textsf{\scriptsize Setup}}(\lambda)$ \mbox{\color{gray}\it\  // For each data source $j$}
\[
S_Y\leftarrow{\textsf{\scriptsize Commit}}({\tt srs},\hat{s}[1,Y]),  \ K\leftarrow{\textsf{\scriptsize Commit}}({\tt srs},\hat{k}[Y])
\]

\item 
${\rm Data}_j \Rightarrow {\rm Prover}$:
\[
D_j\leftarrow{\textsf{\scriptsize Commit}}({\tt srs}_j,d_j[X]),\  \sigma_j\leftarrow{\textsf{\scriptsize Sign}}({\tt sk}_j,D_j)
\]

\item 
{\rm Prover} $\Rightarrow$ {\rm Verifier}:
\[
(D_j, \sigma_j)_{j=1}^J
\]
\[
R\leftarrow{\textsf{\scriptsize Commit}}({\tt srs},r[X,1]), \ \tilde{R}\leftarrow{\textsf{\scriptsize Commit}}({\tt srs},\tilde{r}[X,1])
\]
\item
{\rm Verifier} $\Rightarrow$ {\rm Prover}: $y\xleftarrow{\$}\mathbb {\mathbb F}_p$ {\color{gray}\it// (Fiat-Shamir): $y\leftarrow{\textsf{\scriptsize Hash}}(R|\tilde{R})$}
\item
{\rm Prover} $\Rightarrow$ {\rm Verifier}: 
\[
\begin{array}{r@{}l@{}r@{}l}
T&\leftarrow{\textsf{\scriptsize Commit}}({\tt srs},t[X,y]) \\
S_X&\leftarrow{\textsf{\scriptsize Commit}}({\tt srs},\hat{s}[X,y]) \mbox{\color{gray}\it\  // (Outsourced to Prover)}
\end{array}
\] 
{\scriptsize \ }

\item
{\rm Verifier} $\Rightarrow${\rm Prover}:  $z\xleftarrow{\$}\mathbb {\mathbb F}_p$ {\color{gray}\it// (Fiat-Shamir): $z\leftarrow{\textsf{\scriptsize Hash}}(R|\tilde{R}|T| S_X)$}

\item
{\rm Prover} $\Rightarrow${\rm Verifier}: 
\[
\begin{array}{r@{}l@{}r@{}l}
(r_1, \pi_{r_1})&\leftarrow{\textsf{\scriptsize Open}}({\tt srs},R,z),& \\
(r_2, \pi_{r_2})&\leftarrow{\textsf{\scriptsize Open}}({\tt srs},R,zy) \\ 
(t, \pi_t)&\leftarrow{\textsf{\scriptsize Open}}({\tt srs},T,z), &
 \\
(\tilde{r}, \pi_{\tilde{r}})&\leftarrow{\textsf{\scriptsize Open}}({\tt srs},\tilde{R},z),& \\
(d_j, \pi_{d_j})&\leftarrow{\textsf{\scriptsize Open}}({\tt srs}_j,D_j,z), 1 \le j \le J
\end{array}
\]
{\color{gray}\it// (Outsourced to Prover): }
\[
\left\{
\begin{array}{r@{}l@{\ }r@{}l}
(k, \pi_k)&\leftarrow{\textsf{\scriptsize Open}}({\tt srs},K,y), &
(\hat{s}, \pi_{\hat{s}})&\leftarrow{\textsf{\scriptsize Open}}({\tt srs},S_X,z) \\ 
(\hat{s}_1, \pi_{\hat{s}_1})&\leftarrow{\textsf{\scriptsize Open}}({\tt srs},S_X,1), &
(\hat{s}_2, \pi_{\hat{s}_2})&\leftarrow{\textsf{\scriptsize Open}}({\tt srs},S_Y,y)  
\end{array}
\right.
\]
{\scriptsize \ }

\item
Verifier checks:

{\color{gray}\it// Additionally verify: $r[z,1] \stackrel{?}{=} \tilde{r}[z,1] + \sum_{j=1}^J d_j[z] z^{N + \sum_{j=1}^{J-1} m _j}$}\\
{\color{gray}\it// and the respective polynomial commitment proofs}
\[
\begin{array}{r@{}l}
 &\bigwedge_{j=1}^J   {\textsf{\scriptsize VerifySign}}({\tt pk}_j,D_j,\sigma_j)  \wedge  \\
 &\bigwedge_{j=1}^J  {\textsf{\scriptsize Verify}}({\tt srs}_j,D_j,z,d_j,\pi_{d_j})  \wedge \\
 &{\textsf{\scriptsize Verify}}({\tt srs},\tilde{R},z,\tilde{r},\pi_{\tilde{r}}) \wedge \\
 &\big(r_1 \stackrel{?}{=} \tilde{r} + \sum_{j=1}^J d_j z^{N + \sum_{j=1}^{J-1} m _j}\big)  \wedge\\
 &\big(t \stackrel{?}{=} r_1(r_2+\hat{s})-k\big) \wedge \\
 &(\hat{s}_1 \stackrel{?}{=} \hat{s}_2) \wedge {\textsf{\scriptsize Verify}}({\tt srs},T,z,t,\pi_t) \wedge \\
 &{\textsf{\scriptsize Verify}}({\tt srs},R,z,r_1,\pi_{r_1})  \wedge {\textsf{\scriptsize Verify}}({\tt srs},R,zy,r_2,\pi_{r_2}) \wedge \\
 &{\textsf{\scriptsize Verify}}({\tt srs},K,y,k,\pi_k)  \wedge {\textsf{\scriptsize Verify}}({\tt srs},S_X,z,\hat{s},\pi_{\hat{s}}) \wedge \\
 &{\textsf{\scriptsize Verify}}({\tt srs},S_X,1,\hat{s}_1,\pi_{\hat{s}_1})  \wedge {\textsf{\scriptsize Verify}}({\tt srs},S_Y,y,\hat{s}_2,\pi_{\hat{s}_2})
\end{array}
\]

\end{enumerate}    
\end{longfbox} \vspace{-5pt}
\end{table}

\section{Application to Parametric Bushfire Insurance} \label{sec:model}

In this section, we apply the zk-SNARK protocol to develop a framework for private-preserving parametric insurance. We aim to satisfy the following privacy and security requirements:
\begin{itemize}

    \item \textbf{Private Data Concealment}: An insuree's private data (e.g., the insured location) should not be revealed to any other users, except from the insurer and data sources.

    \item \textbf{False Claim Prevention}: We assume no trust on the insuree. That is, the insuree may be dishonest and try to claim the insurance even though the claim conditions are not satisfied. Apart from completeness, we also require knowledge soundness, such that it is impossible for a dishonest prover to claim the insurance with false data. 

    \item \textbf{Efficient On-chain Verification}: Claiming the insurance should be efficient and low cost. The computation for parametric index may be complex, but we require efficient on-chain claim verification. 
	
\end{itemize}

\subsection{Remote Sensing Model for Bushfire Detection}\label{sec:sensing}

\begin{figure}[t]  
	\centering
	  \includegraphics[width=\linewidth]{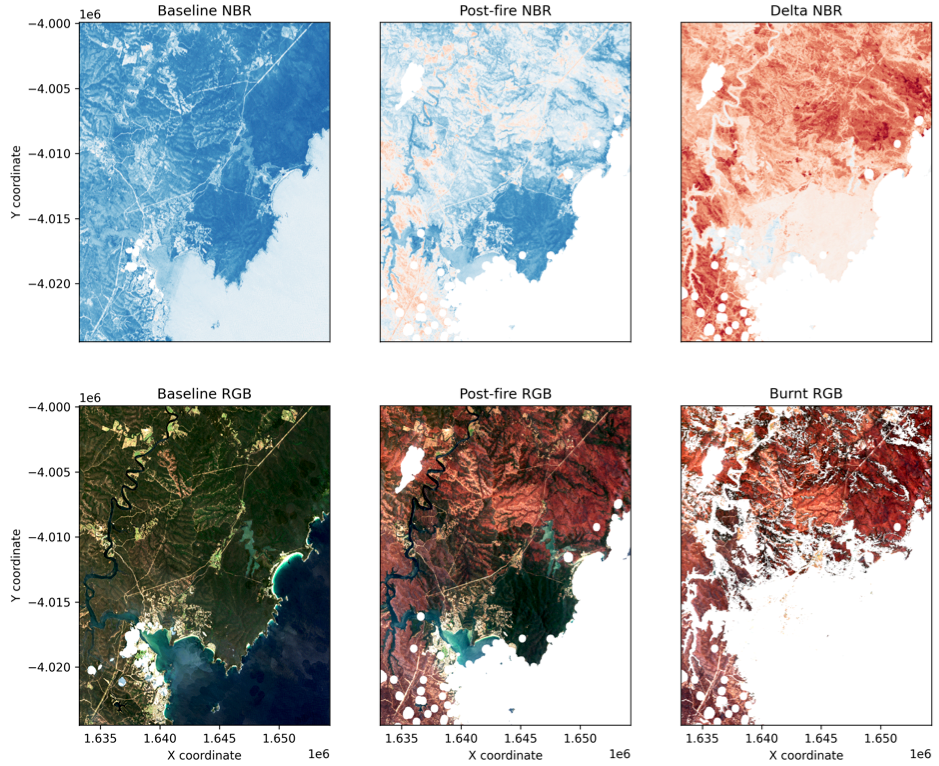} 
	  \caption{Example satellite images from Sentinel-2B MSI Definitive ARD dataset. The top row shows the NBR before and after the bushfire and the resulting dNBR. The bottom row shows the original satellite images and the burnt areas that are masked according to the dNBR threshold 0.3.}
	  \label{fig:nbr}
\end{figure}

We demonstrate the application of our protocol by incorporating a bushfire detection model into our insurance claim handling process. While there are several models for estimating bushfire severity from satellite imaginary \cite{he2011numerical, keeley2009fire}, a common model is the delta Normalized Burn Ratio (dNBR) \cite{keeley2009fire}, which is used for detecting bushfire in this paper. The dNBR is based on the difference between ground electromagnetic waves of normal areas and burnt areas. In general, a normal area has very high reflectance in the Near Infrared Spectral Regions (NIR) and low reflectance in the Shortwave Infrared Spectral Regions (SWIR). In contrast, in a burnt area, the NIR is much lower than SWIR \cite{keeley2009fire}. Define the Normalized Burn Ratio (NBR) as the proportional difference between the two spectral regions by:
\[
    {\rm NBR} \triangleq \frac{{\rm NIR}-{\rm SWIR}}{{\rm NIR}+{\rm SWIR}}
\]
If an area is severely damaged by bushfire, the difference on NBR before and after the fire is high. Thus, define the burnt severity index (dNBR) by:
\[
    {\rm dNBR} \triangleq {\rm NBR}_{\rm prefire}-{\rm NBR}_{\rm postfire}
\]
According to the United States Geological Survey \cite{van2013climatic}, dNBR with a value larger than 0.66 is considered high severity. NIR and SWIR can be obtained from satellite imagery via remote sensing. For example, Digital Earth Australia (DEA) provides satellite imaginary datasets over Australia with very high precision \cite{Krause_Dunn_Bishop-Taylor_2021}. See some examples in Fig~\ref{fig:nbr}.

Consider a ground area represented by $n$ pixels in a satellite image (each pixel can cover an area of 20-50 m$^2$). We define the overall burnt severity over an area as the proportion of pixels with high {\rm dNBR}. For each $i$-th pixel, let the pre-fire NIR, SWIR, NBR be $\mathbf r^-_i$, $\mathbf s^-_i$ and $\mathbf n^-_i$, and the post-fire NIR, SWIR, NBR be $\mathbf r^+_i$, $\mathbf s^+_i$ and $\mathbf n^+_i$. The pixels having ${\rm dNBR}>\kappa$ are considered severely burnt. If the total number of severely-burnt pixels in a nearby area is more than a threshold $\epsilon$, then the area is considered severely burnt. The constraint system of a bushfire insurance claim can be formulated as follows:
\[
\left\{
\begin{array}{l@{\ }ll}
(\mathbf r^-_i - \mathbf s^-_i) = \mathbf{n}^-_i (\mathbf r^-_i + \mathbf s^-_i)+\mathbf \theta^-_i,& \mbox{for\ } 1 \le i \le n \\
(\mathbf r^+_i - \mathbf s^+_i) = \mathbf{n}^+_i (\mathbf r^+_i + \mathbf s^+_i)+\mathbf \theta^+_i,& \mbox{for\ } 1 \le i \le n \\
\theta_{\max} - \sum^n_{i=1} (\mathbf \theta^-_i)^2 - \sum^n_{i=1} (\mathbf \theta^+_i)^2 = \theta_{d} > 0 \\
G = \sum^{n}_{i=1} {\mathds 1} ( \mathbf{n}^-_i - \mathbf{n}^+_i \ge \kappa ) - \epsilon 
\end{array}
\right.
\]
where $\theta$ is the rounding error of integer division with a tolerance of sum square error $\theta_{\max}$, and ${\mathds 1}(\cdot)$ be an indicator function. A bushfire insurance claim will be valid, if $G\ge0$.

The constraint ($G = \sum^{n}_{i=1} {\mathds 1}( \mathbf{n}^-_i - \mathbf{n}^+_i \ge \kappa) - \epsilon$) can be re-expressed as a set of multiplicative and linear constraint equations as follows:
\[
\left\{
\begin{array}{@{}r@{\ }ll}
\mathbf{i}_i (1 - \mathbf{i}_i) =  0,& \mbox{for\ } 1 \le i \le n \\
\mathbf{e}_{i,j} (\mathbf{e}_{i,j} - 2^{j-1} ) = 0,& \mbox{for\ } 1 \le i \le n,  1 \le j \le k \\
\sum_{j=1}^k \mathbf{e}_{i,j} - \mathbf{i}_i (\mathbf{n}^-_i - \mathbf{n}^+_i  - \kappa) = 0,& \mbox{for\ } 1 \le i \le n \\
G + \epsilon \phantom{1} = \sum^{n}_{i=1} \mathbf{i}_i &
\end{array} 
\right.
\]
where $\mathbf{i}_i$ is an indicator whether the $i$-th pixel having ${\rm dNBR}>\kappa$, and $(\mathbf{e}_{i,j})_{j=1}^k$ is the $k$-bit-decomposition of the $i$-th pixel's $({\rm dNBR} - \kappa)$, if it is non-negative. If $({\rm dNBR} - \kappa)$ is negative, then $\mathbf{e}_{i,j}$ needs to be zero to make the constraint satisfiable. 

Evidently, the above constraint system can be proved by Sonic zk-SNARK protocol \cite{maller2019sonic}. We specify the settings of Sonic-specific vectors $({\bf a}, {\bf b}, {\bf c})$ and $({\mathbf u_q, \mathbf  v_q, \mathbf  w_q}, k_q)$ with a more detailed explanation in \cite{HQC24full}. There are in total $7n+(n+2)k+2$ linear constraints, and $10n+(n+2)k$ multiplication constraints, where $n$ is the image size and $k$ is the length of bit indices.

\subsection{Private Location Hiding in Input Data}

In order to ensure the accuracy of insurance claims, it is essential to verify that the satellite images used as input data correspond to the correct insured location. To achieve this, one can request the data source to hash the location of a satellite image as $H = {\sf Hash}({\rm location})$. Note that $H$ in the smart contract does not reveal the true location. The data source then signs on the concatenated message ($H|D_j$) as follows:
\[
\sigma_j\leftarrow{\textsf{\small Sign}}({\tt sk}_j, H|D_j )
\]
To verify the signature $\sigma_j$, the prover provides $D_j$, but $H$ is encoded in the smart contract for on-chain verification as:
\[
 {\textsf{\small VerifySign}}({\tt pk}_j, H|D_j,\sigma_j) 
\]

We remark that our approach can be generally applied to hide any specific private data in the input, while enabling proper verification based on the private data. For example, in the context of blockchain-based flight delay insurance, an insuree needs to provide a zero-knowledge proof of flight delay information to prove that the flight number and date match with the required ones in the insurance policy. In this scenario, the data source can hash the flight number and date and sign the combined hash and commitment value. This ensures that the hashed part will be verified on blockchain without revealing the actual flight number and date.
\section{Implementation and Evaluation} \label{sec:eval}

This section presents an evaluation study of our protocol on real-world permissionless blockchain platform Ethereum \cite{dannen2017introducing}. We implemented the parametric bushfire insurance application in Sec.~\ref{sec:model} and the on-chain verification protocol in Sec.~\ref{sec:extensions} as smart contracts by Solidity programming language.

{\bf Data:}
To prepare the dataset of satellite images for bushfire insurance claims, we selected 57 locations, ranging from the vicinity of Brisbane to the southern coast of Australia, which have been severely affected by bushfire in 2019. For each location, two snapshots were chosen: Jul 2019 and Feb 2020, which are before and after the bushfire season. For each snapshot, both NIR and SWIR images were acquired from the DEA satellite data repository \cite{Krause_Dunn_Bishop-Taylor_2021}.

{\bf Evaluation Environments:}
The evaluation of the prover was conducted on a Google Cloud with virtual machine E2 series 16v CPU. The evaluation of verification was conducted on Goerli (a testnet of Ethereum \cite{moralis2022goerli}). We repeated each experiment at least 10 times to obtain average measurements.

{\bf Verifier Implementation on Smart Contracts:}
We divide the on-chain verification of an insurance claim into two smart contracts: (1) {\em individual policy contract} that deals with individual requirements (e.g., insurance policy for a specific location with private location hiding), (2) {\em global policy contract} that deals with general validity requirements (e.g., the criteria for severely burnt areas). Note that the global policy contract is deployed only once for all insurees.

We consider different versions of verifiers on smart contract:
\begin{enumerate}
\item \textbf{\em Sonic Verifier}: The verifier of the Sonic protocol (${\tt S}_{\tt 0}$) in Table~\ref{tab:datasource}.

\item \textbf{\em Enhanced Verifier}: The verifier of enhanced protocol (${\tt S}_{\tt EV}$) with batch verification (see \cite{HQC24full}).

\item \textbf{\em Enhanced+ Verifier}: An improved verifier of enhanced protocol with off-chain SRS, in which the necessary SRS elements are stored in a trusted off-chain party (e.g., blockchain oracles) and are only sent to the verifier through an oracle when requested, thereby further reducing gas costs. 

\end{enumerate}

\vspace{-5pt}
\subsection{Performance of Protocol}

We measure our performance in the following aspects: 
\begin{enumerate}
\item \textbf{Input Size}: The number of pixels of a satellite image.
\item \textbf{Num. Linear Constraints}: The number of linear constraints included in the constraint system. 
\item \textbf{Num. Multiplication Constraints}: The number of multiplication constraints in the constraint system. 
\item \textbf{SRS}: The size of the Structured Reference String (SRS) in MB and the running time to generate the SRS in minutes. A universal SRS was utilized for all experiments. Thus, the SRS size remains constant.
\item \textbf{Verifier SRS}: The size of the SRS elements that are needed for on-chain verification. This is measured using the number of uint256 (Solidity variable type).
\item \textbf{Prover Memory Space}: The size of the generated polynomials during the proof generation, which includes $S_Y, K, D_j, R, T$, and $S_X$. Note that these polynomials are only stored in the prover  during proof generation and will be subsequently deleted.
\item \textbf{Proving Time}: The time of proof generation for a claim.
\item \textbf{Proof Size}: The size of a proof for on-chain verification. A proof contains 39 uint256 variables (each has 32 bytes).
\item \textbf{Verification Time}: The time required for on-chain verification on the testnet of Ethereum.
\end{enumerate}

\begin{table}[t]
    \centering
    \begin{threeparttable}
    \caption{Performance Evaluation Results\tnote{$\dag$}} \vspace{-5pt}
    \begin{tabular}{@{}c||c|c|c|c|c@{}}
    \hline
    Input Size (pixels) & 4 & 8 & 16 & 32 & 64 \\
    \hhline{|=||=|=|=|=|=|}
    Num. Linear Cons. & 232 & 400 & 736 & 1408 & 2752 \\
    \hline
    Num. Multiply Cons. & 222 & 378 & 690 & 1314 & 2562 \\
    \hline
    SRS (MB) & 191.6 & 191.6 & 191.6 & 191.6 & 191.6 \\
    \hline
    SRS (min) & 801 & 801 & 801 & 801 & 801\\
    \hline
    Verifer SRS (\# Elements) & 14 & 14 & 14 & 14 & 14 \\
    \hline
    Prover Memory Space (MB) & 16.9 & 51.5 & 178.3 & 659.3 & 6510 \\
    \hline
    Proving Time (sec) & 177 & 350 & 652 & 1615 & 5061 \\
    \hline
    Proof Size (KB) & 1.22 & 1.22 & 1.22 & 1.22 & 1.22 \\
    \hline
    Verification Time (sec) & 7.09 & 6.98 & 7.14 & 7.07 & 7.15 \\
    \hline
    \end{tabular}
    \label{tab:exp_results}
    \begin{tablenotes}
        \footnotesize
        \item[$\dag$] 
        Evaluation based on the implementation of the protocol in Table \ref{tab:datasource} using the commitment scheme in Table \ref{tab:batchkzg}.
      \end{tablenotes}
    \end{threeparttable} \vspace{-5pt}
\end{table}

\begin{figure*}[t]  
\centering
	\begin{minipage}{0.28\textwidth}
    \hspace{-10pt}
	  \includegraphics[width=1\linewidth]{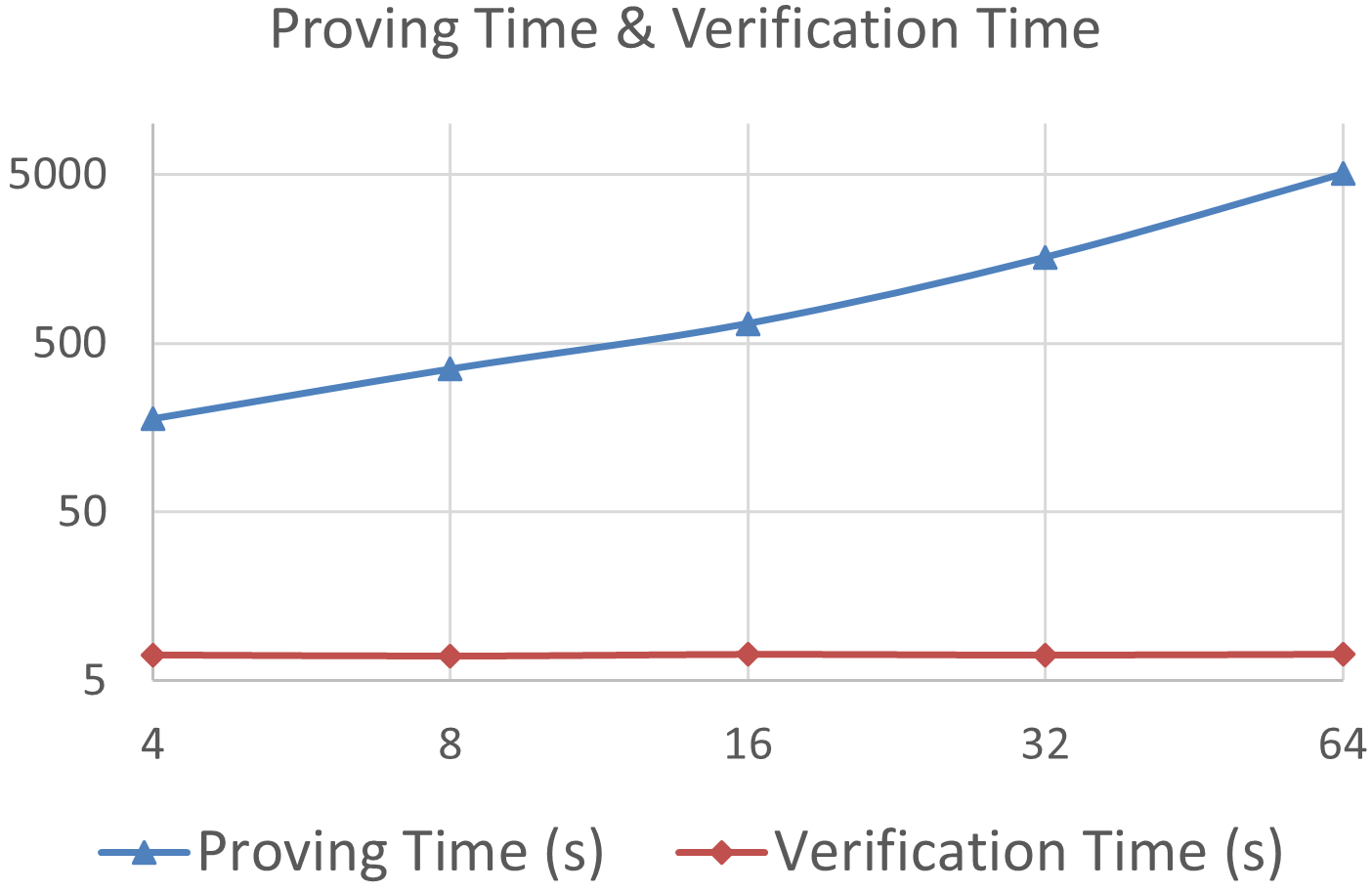} 
	  \caption{Proving time \& verification time\\ in Table \ref{tab:exp_results}.}
	  \label{fig:exp_results}
	 \end{minipage} 
	\begin{minipage}{0.35\textwidth}
    \hspace{-8pt}  
	  \includegraphics[width=1.05\linewidth]{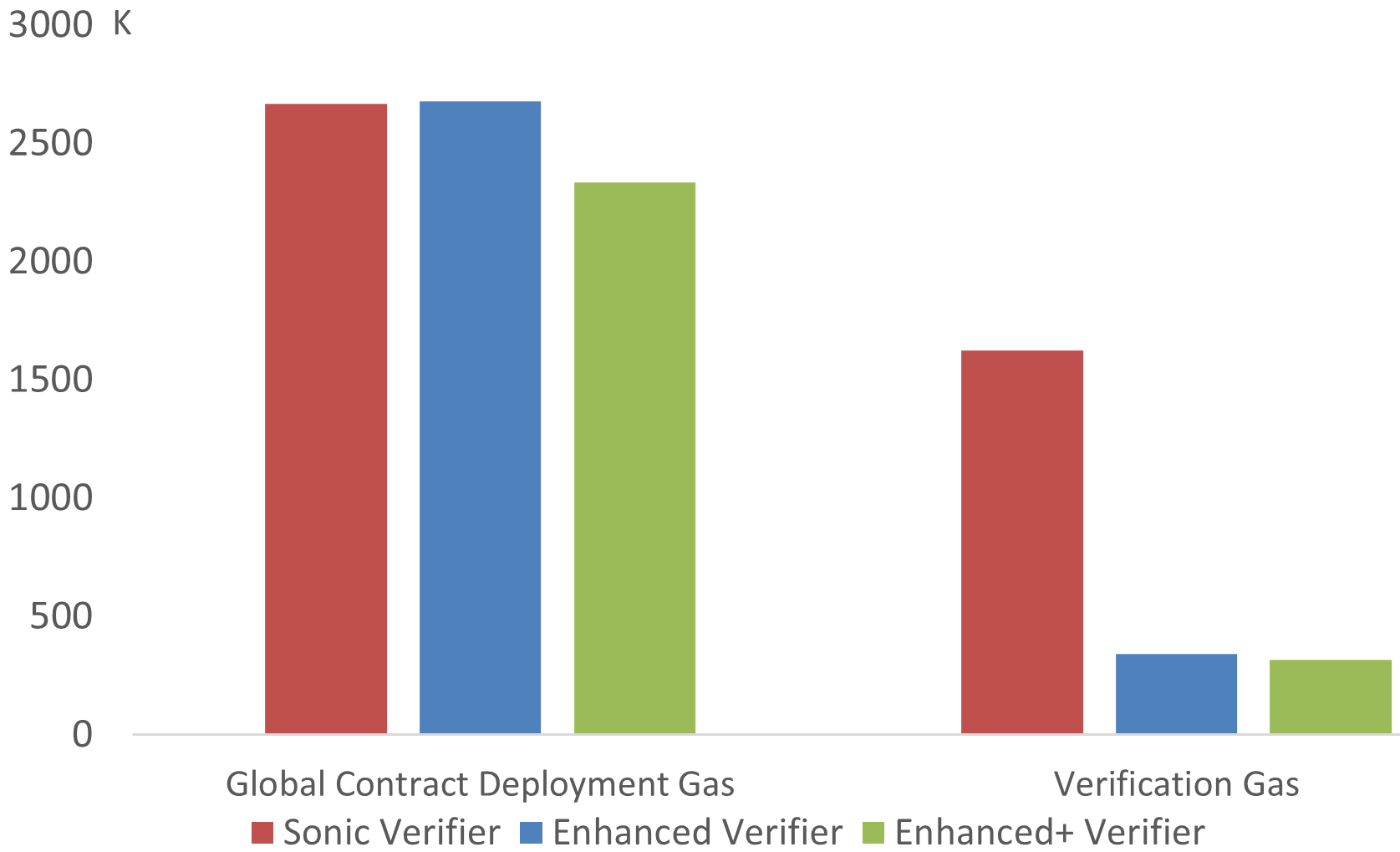} \vspace{-20pt}
	  \caption{A comparison of verifiers in Table \ref{tab:cost_results}.}
	  \label{fig:gas cost bar chart}
	 \end{minipage} \	 
	\begin{minipage}{0.28\textwidth}
    \centering
	  \includegraphics[width=1\linewidth]{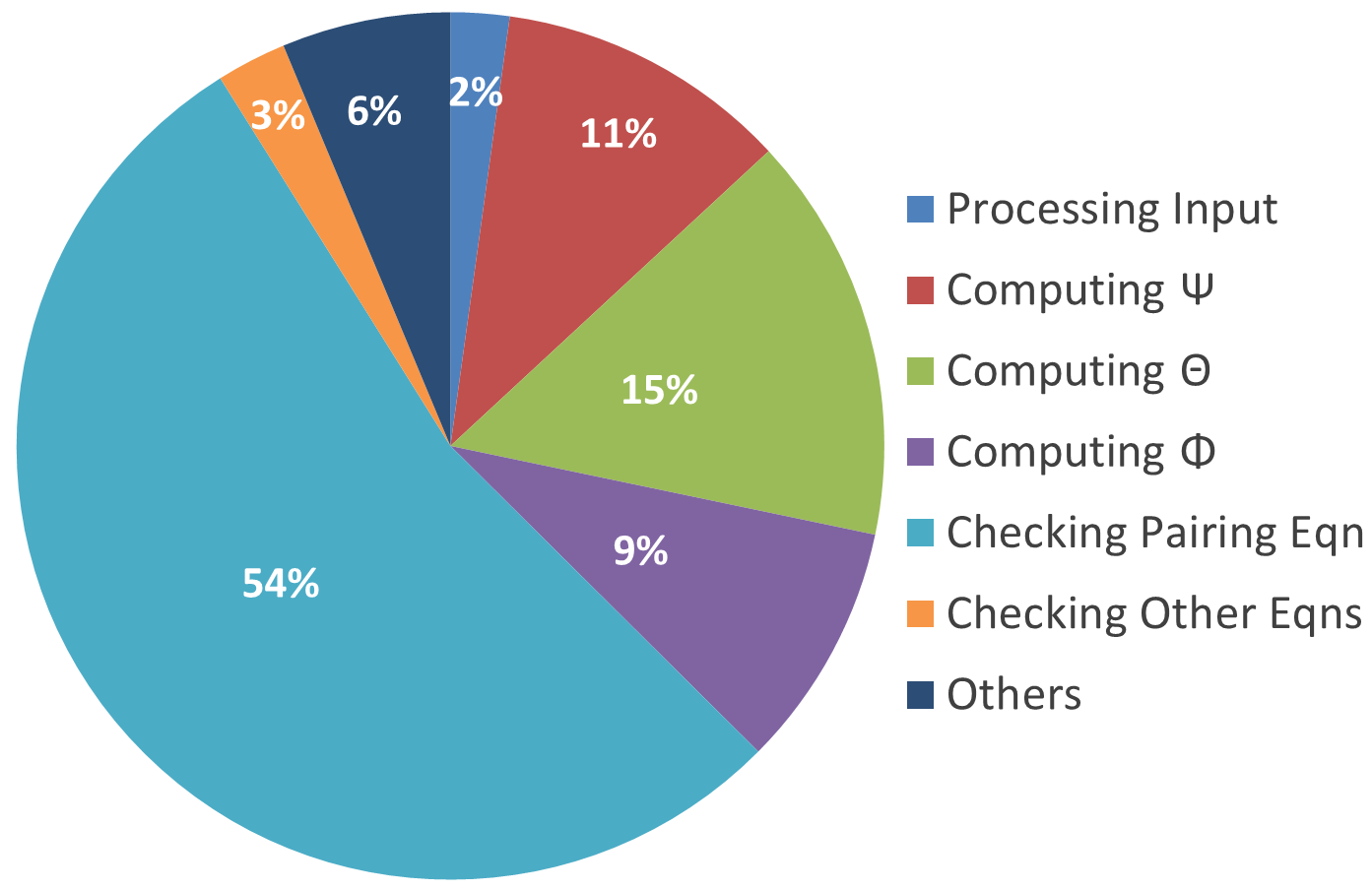}  
	  \caption{Breakdown of enhanced verifier gas cost in Table \ref{tab:batchdistri}.}
	  \label{fig:gas cost pie chart} \vspace{-5pt}
	 \end{minipage}
\end{figure*}

We present the evaluation results for enhanced verifier in Table~\ref{tab:exp_results}. We observe that the numbers of linear and multiplication constraints grow linearly with the input size, so does the proving time, because of correlation. For an input size of 64 pixels, a notable increase in proving time is observed from 1615 sec for input sizes of 32 pixels, to 5061 sec. This is due to insufficient memory in our test machine, which can be improved with larger memory.
In Fig.~\ref{fig:exp_results}, the proving time grows linearly, and the proof memory space grows quadratically with respect to the input size. In contrast, the verification time remains constant regardless of the input size, since the proof size remains constant relative to the input size. Although the size of the SRS is significant, only a fixed number of elements are required for verification. 
Note that we did not consider multi-core processor optimization, which will be explored in future work.

\begin{table}[t]
    \centering
    \begin{threeparttable}
    \caption{Gas Costs of Different Verifiers} \vspace{-5pt}
    \begin{tabular}{@{}cc|c|c|c}
    \hline
    \multicolumn{2}{@{}c|}{}  & Sonic & Enhanced  & Enhanced+  \\ 
    \multicolumn{2}{@{}c|}{}  & Verifier & Verifier & Verifier \\ \hline \hline
    \multicolumn{1}{@{}c@{}|}{\multirow{3}{*}{\begin{tabular}[c]{@{}c@{}}Global Contract\\ Deployment\\ Cost\end{tabular}}}                         & Gas  &  2667K    & 2677K & 2334K           \\ \cline{2-5} 
    \multicolumn{1}{@{}c|}{} & ETH  &  0.05334    & 0.05355 &  0.04668          \\ \cline{2-5} 
    \multicolumn{1}{@{}c|}{} & USD\$ &  63.97    & 64.21 & 55.98           \\ \cline{2-5} \hline
   \multicolumn{1}{@{}c@{}|}{\multirow{3}{*}{\begin{tabular}[c]{@{}c@{}}Individual Contract\\ Deployment\\ Cost\end{tabular}}}                         & Gas  &  156K    & 156K & 156K           \\ \cline{2-5} 
    \multicolumn{1}{@{}c|}{} & ETH  &  0.0031    & 0.0031 &  0.0031          \\ \cline{2-5} 
    \multicolumn{1}{@{}c|}{} & USD\$ &  3.73    & 3.73 & 3.73           \\ \cline{2-5} \hline	
    \multicolumn{1}{@{}c@{}|}{\multirow{3}{*}{\begin{tabular}[c]{@{}c@{}}Verification\\ Cost\end{tabular}}}                         & Gas  &  1622K   & 341K & 315K           \\ \cline{2-5} 
    \multicolumn{1}{@{}c|}{} & ETH  & 0.03245     & 0.00683 & 0.0063           \\ \cline{2-5} 
    \multicolumn{1}{@{}c|}{} & USD\$ & 38.91     & 8.19 & 7.56           \\ \cline{2-5} \hline
    \end{tabular}
    \label{tab:cost_results}
    \end{threeparttable} \vspace{-5pt}
\end{table}

\vspace{-5pt}
\subsection{Comparison of Gas Costs of Different Verifiers}

We present the gas costs of different verifiers in Table~\ref{tab:cost_results}. The gas costs are estimated based on ETH/USD\$ = 1199.11 (as of the end of 2022 \cite{ycharts2023crypto}) and the average gas price as 20 gwei. We observe that the gas costs incurred by global contract deployment are around seven times greater than the ones of on-chain verification. However, the most deployment gas costs can be amortized, because the global policy contract is shared among all the users. The individual policy contract deployment costs only 5\% of the one of global contract. We note that enhanced verifier can significantly reduce the verification cost by around 78\%. Notably, the cost of enhanced verifier can be further optimized by storing SRS off-chain through a trusted third party. In this way, necessary SRS elements are only sent to the verifier along with a proof, instead of being stored on-chain. As a result, we observe a greater reduction by enhanced and enhanced+ verifiers on verification gas cost by 78\% and 80\%, respectively, compared to Sonic verifier in Fig.~\ref{fig:gas cost bar chart}. Therefore, our results significantly enhance the practicality of zk-SNARKs to real-world blockchain-enabled applications.

\begin{table}[t]
    \centering
    \begin{threeparttable}
    \caption{Breakdown of Gas Cost of Sonic Verifier} \vspace{-5pt}
    \begin{tabular}{l|c|c|c}
\hline
\multicolumn{1}{c|}{Operations} & \multicolumn{1}{c|}{Gas} & ETH & USD\$ \\ \hline \hline
Processing Input                                             & 3K                              & 0.0004       & 0.51       \\ \hline
Checking 9 Pairing Equations                              & 1590K                            & 0.0318       & 38.13       \\ \hline
Checking Other Equations                                          & 7K                              & 0.0001       & 0.17       \\ \hline
Others                                                         & 22K                             & 0.0004       & 0.51       \\ \hline \hline
Total                                                          & 1622K                            & 0.0324       & 38.91       \\ \hline
\end{tabular}
    \label{tab:sonicdistri}
    \end{threeparttable} \vspace{-5pt}
\end{table}

\begin{table}[t]
    \centering
    \begin{threeparttable}
    \caption{Breakdown of Gas Cost of Enhanced Verifier} \vspace{-5pt}
    \begin{tabular}{l|c|c|c}
\hline
\multicolumn{1}{c|}{Operations} & \multicolumn{1}{c|}{Gas} & ETH & USD\$ \\ \hline \hline
Processing Input                                             & 7K                              & 0.0001       & 0.18       \\ \hline
Computing $\Psi_i$                               & 37K                             & 0.0007       & 0.89       \\ \hline
Computing $\Theta$                                                    & 52K                             & 0.0010       & 1.24       \\ \hline
Computing $\Phi$                                                   & 31K                             & 0.0006       & 0.75       \\ \hline
Checking Single Pairing Equation                              & 182K                            & 0.0036       & 4.38       \\ \hline
Checking Other Equations                                          & 9K                              & 0.0002       & 0.21       \\ \hline
Others                                                         & 22K                             & 0.0004       & 0.51       \\ \hline \hline
Total                                                          & 341K                            & 0.0068       & 8.19       \\ \hline
\end{tabular}
    \label{tab:batchdistri}

    \begin{tablenotes}
        \footnotesize
        \item[$\dag$]
        $\Psi_i, \Theta, \Phi$ are defined in ${\textsf{\scriptsize BatchVerify}}_{\tt rKZGb}$ in Table \ref{tab:batchdistri}. \end{tablenotes}
    \end{threeparttable} \vspace{-5pt}
\end{table}

\vspace{-5pt}
\subsection{Breakdowns of Verifier Gas Costs}

In this subsection, we further break down the verification gas costs for Sonic verifier and enhanced verifier. The breakdowns of gas costs are listed in Table~\ref{tab:batchdistri}, Table~\ref{tab:sonicdistri} and Fig.~\ref{fig:gas cost pie chart}. We observe that Sonic verifier costs over 98\% of gas on checking the 9 pairing equations. Although we adopted a pre-compiled contract \cite{eip197} on Ethereum for checking pairing equations, checking each pairing equation is still costly. On the contrary, our batch polynomial commitment scheme can dramatically reduce the number of pairing equations into a single equation, reducing to only 11\% of the previous gas cost. Hence, this makes a substantial reduction in the gas costs of verifying an insurance claim with only US\$8.19.

\section{Conclusion}

In this paper, we proposed a privacy-preserving parametric insurance framework based on blockchain to offer an efficient privacy-preserving solution for insurance policies. The use of zk-SNARKs allows for private verification of insurance claims and data authenticity, reducing the risk of fraudulent activities and maintaining user privacy. Our proof-of-concept of bushfire parametric insurance on the Ethereum blockchain has demonstrated the effectiveness of our framework.

In future work, we will apply the enhanced Sonic zk-SNARK framework to a wide range of privacy-preserving blockchain-enabled applications \cite{LCWZ22sharing,CZ22energyplan,ZC21energyplan,WCZ21energyshare,WCZ22energyshare,ZCGL22tiot}.

\bibliographystyle{IEEEtran}
\bibliography{bib}

\begin{thebibliography}{10}
\providecommand{\url}[1]{#1}
\csname url@samestyle\endcsname
\providecommand{\newblock}{\relax}
\providecommand{\bibinfo}[2]{#2}
\providecommand{\BIBentrySTDinterwordspacing}{\spaceskip=0pt\relax}
\providecommand{\BIBentryALTinterwordstretchfactor}{4}
\providecommand{\BIBentryALTinterwordspacing}{\spaceskip=\fontdimen2\font plus
\BIBentryALTinterwordstretchfactor\fontdimen3\font minus \fontdimen4\font\relax}
\providecommand{\BIBforeignlanguage}[2]{{%
\expandafter\ifx\csname l@#1\endcsname\relax
\typeout{** WARNING: IEEEtran.bst: No hyphenation pattern has been}%
\typeout{** loaded for the language `#1'. Using the pattern for}%
\typeout{** the default language instead.}%
\else
\language=\csname l@#1\endcsname
\fi
#2}}
\providecommand{\BIBdecl}{\relax}
\BIBdecl

\bibitem{lin2020application}
X.~Lin and W.~J. Kwon, ``Application of parametric insurance in principle-compliant and innovative ways,'' \emph{Risk Management and Insurance Review}, vol.~23, no.~2, pp. 121--150, 2020.

\bibitem{insuresilience}
MRPP2020, ``{Global parametrics - Mexican reef protection program}.''

\bibitem{mingyu2023blochchain}
M.~Hao, K.~Qian, and S.~C.-K. Chau, ``Blockchain-enabled parametric solar energy insurance via remote sensing,'' in \emph{ACM Intl. Conf. on Future Energy Systems (e-Energy)}, 2023.

\bibitem{Etherisc}
\BIBentryALTinterwordspacing
Etherisc, ``Etherisc,'' Apr 2022. [Online]. Available: \url{https://etherisc.com}
\BIBentrySTDinterwordspacing

\bibitem{biryukov2019deanonymization}
A.~Biryukov and S.~Tikhomirov, ``Deanonymization and linkability of cryptocurrency transactions based on network analysis,'' in \emph{IEEE European symposium on security and privacy (EuroS\&P)}, 2019.

\bibitem{gasandfees}
\BIBentryALTinterwordspacing
Ethreum.org, ``Gas and fees,'' Feb 2023. [Online]. Available: \url{https://ethereum.org/en/developers/docs/gas/}
\BIBentrySTDinterwordspacing

\bibitem{rackoff1991non}
C.~Rackoff and D.~R. Simon, ``Non-interactive zero-knowledge proof of knowledge and chosen ciphertext attack,'' in \emph{{Annual International Cryptology Conference}}, 1991, pp. 433--444.

\bibitem{maller2019sonic}
M.~Maller, S.~Bowe, M.~Kohlweiss, and S.~Meiklejohn, ``Sonic: Zero-knowledge snarks from linear-size universal and updatable structured reference strings,'' in \emph{Proc. the ACM SIGSAC Conf. Computer and Communications Security (CCS)}, 2019, pp. 2111--2128.

\bibitem{b3i}
B3i, ``{Munich re Blockchain initiative B3i gains truly international scope},'' Feb 2017.

\bibitem{he2018toward}
X.~He, S.~Alqahtani, and R.~Gamble, ``Toward privacy-assured health insurance claims,'' in \emph{Intl. Conf. on Internet of Things}, 2018, pp. 1634--1641.

\bibitem{cormode2012practical}
G.~Cormode, M.~Mitzenmacher, and J.~Thaler, ``Practical verified computation with streaming interactive proofs,'' in \emph{Proc. Innovations in Theoretical Computer Science Conference}, 2012, pp. 90--112.

\bibitem{blumberg2014verifiable}
A.~J. Blumberg, J.~Thaler, V.~Vu, and M.~Walfish, ``Verifiable computation using multiple provers,'' \emph{Cryptology ePrint Archive}, 2014.

\bibitem{Flashproofs}
\BIBentryALTinterwordspacing
N.~Wang and S.~C.-K. Chau, ``Flashproofs: Efficient zero-knowledge arguments of range and polynomial evaluation with transparent setup,'' in \emph{{IACR AsiaCrypt}}, 2022. [Online]. Available: \url{https://eprint.iacr.org/2022/1251}
\BIBentrySTDinterwordspacing

\bibitem{thaler2022proofs}
J.~Thaler, ``Proofs, arguments, and zero-knowledge,'' 2022.

\bibitem{parno2016pinocchio}
B.~Parno, J.~Howell, C.~Gentry, and M.~Raykova, ``Pinocchio: Nearly practical verifiable computation,'' \emph{Communications of the ACM}, vol.~59, no.~2, pp. 103--112, 2016.

\bibitem{gabizon2019plonk}
A.~Gabizon, Z.~J. Williamson, and O.~Ciobotaru, ``Plonk: Permutations over lagrange-bases for oecumenical noninteractive arguments of knowledge,'' \emph{Cryptology ePrint Archive}, 2019.

\bibitem{ben2018fast}
E.~Ben-Sasson, I.~Bentov, Y.~Horesh, and M.~Riabzev, ``Fast reed-solomon interactive oracle proofs of proximity,'' in \emph{International Colloquium on Automata, Languages, and Programming (ICALP)}, 2018.

\bibitem{boneh21kzg}
D.~Boneh, J.~Drake, B.~Fisch, and A.~Gabizon, ``Efficient polynomial commitment schemes for multiple points and polynomials,'' \emph{Cryptology ePrint Archive}, 2021.

\bibitem{groth2016size}
J.~Groth, ``On the size of pairing-based non-interactive arguments,'' in \emph{{Annual Intl. Conf. on Theory and Applications of Cryptographic Techniques}}, 2016, pp. 305--326.

\bibitem{chiesa2020marlin}
A.~Chiesa, Y.~Hu, M.~Maller, P.~Mishra, N.~Vesely, and N.~Ward, ``Marlin: preprocessing zksnarks with universal and updatable srs,'' in \emph{{Annual Intl. Conf. on the Theory and Applications of Cryptographic Techniques}}, 2020, pp. 738--768.

\bibitem{bunz2020transparent}
B.~B{\"u}nz, B.~Fisch, and A.~Szepieniec, ``Transparent snarks from dark compilers,'' in \emph{Advances in Cryptology--EUROCRYPT 2020: 39th Annual Intl. Conf. on the Theory and Applications of Cryptographic Techniques, Zagreb, Croatia, May 10--14, 2020, Proceedings, Part I 39}.\hskip 1em plus 0.5em minus 0.4em\relax Springer, 2020, pp. 677--706.

\bibitem{ben2018scalable}
E.~Ben-Sasson, I.~Bentov, Y.~Horesh, and M.~Riabzev, ``Scalable, transparent, and post-quantum secure computational integrity,'' \emph{Cryptology ePrint Archive}, 2018.

\bibitem{lee2021dory}
J.~Lee, ``Dory: Efficient, transparent arguments for generalised inner products and polynomial commitments,'' in \emph{Intl. Conf. on Theory of Cryptography}, 2021, pp. 1--34.

\bibitem{crytobk}
W.~J. Buchanan, \emph{Cryptography}.\hskip 1em plus 0.5em minus 0.4em\relax River Publishers, 2017.

\bibitem{kate2010constant}
A.~Kate, G.~M. Zaverucha, and I.~Goldberg, ``Constant-size commitments to polynomials and their applications,'' in \emph{Intl. Conf. on Theory and application of Cryptology and Information Security}, 2010, pp. 177--194.

\bibitem{eip4844}
\BIBentryALTinterwordspacing
{Ethereum Community}, ``{EIP-4844},'' 2022. [Online]. Available: \url{https://www.eip4844.com/}
\BIBentrySTDinterwordspacing

\bibitem{snarkintro}
A.~Nitulescu, ``zk-snarks: A gentle introduction,'' Tech. Rep., 2019.

\bibitem{bootle2016efficient}
J.~Bootle, A.~Cerulli, P.~Chaidos, J.~Groth, and C.~Petit, ``Efficient zero-knowledge arguments for arithmetic circuits in the discrete log setting,'' in \emph{{Annual Intl. Conf. on Theory and Applications of Cryptographic Techniques}}, 2016, pp. 327--357.

\bibitem{HQC24full}
\BIBentryALTinterwordspacing
M.~Hao, K.~Qian, and S.~C.-K. Chau, ``{Privacy-preserving Blockchain-enabled Parametric Insurance via Remote Sensing and IoT},'' 2023. [Online]. Available: \url{https://arxiv.org/abs/2305.08384}
\BIBentrySTDinterwordspacing

\bibitem{he2011numerical}
Y.~He, K.~C. Kwok, G.~Douglas, and I.~Razali, ``Numerical investigation of bushfire-wind interaction and its impact on building structure,'' \emph{Fire Saf. Sci}, vol.~10, pp. 1449--1462, 2011.

\bibitem{keeley2009fire}
J.~E. Keeley, ``Fire intensity, fire severity and burn severity: a brief review and suggested usage,'' \emph{International J. wildland fire}, vol.~18, no.~1, pp. 116--126, 2009.

\bibitem{van2013climatic}
P.~J. van Mantgem, J.~C. Nesmith, M.~Keifer, E.~E. Knapp, A.~Flint, and L.~Flint, ``Climatic stress increases forest fire severity across the western u nited s tates,'' \emph{Ecology letters}, vol.~16, no.~9, pp. 1151--1156, 2013.

\bibitem{Krause_Dunn_Bishop-Taylor_2021}
\BIBentryALTinterwordspacing
C.~Krause, B.~Dunn, and R.~Bishop-Taylor, ``Digital earth australia notebooks and tools repository,'' 2021. [Online]. Available: \url{http://pid.geoscience.gov.au/dataset/ga/145234}
\BIBentrySTDinterwordspacing

\bibitem{dannen2017introducing}
C.~Dannen, \emph{Introducing Ethereum and solidity}.\hskip 1em plus 0.5em minus 0.4em\relax Springer, 2017, vol.~1.

\bibitem{moralis2022goerli}
\BIBentryALTinterwordspacing
Moralis, ``{Goerli ETH – What is the Goerli Testnet?}'' Feb 2022. [Online]. Available: \url{https://moralis.io/goerli-eth-what-is-the-goerli-testnet/}
\BIBentrySTDinterwordspacing

\bibitem{ycharts2023crypto}
\BIBentryALTinterwordspacing
CryptoCompare, ``Ethereum price,'' Feb 2023. [Online]. Available: \url{https://ycharts.com/indicators/ethereum_price}
\BIBentrySTDinterwordspacing

\bibitem{eip197}
\BIBentryALTinterwordspacing
{Ethereum Community}, ``{EIP-197}: Precompiled contracts for optimal pairing check on the elliptic curve alt bn128,'' 2022. [Online]. Available: \url{https://eips.ethereum.org/EIPS/eip-197}
\BIBentrySTDinterwordspacing

\bibitem{LCWZ22sharing}
L.~Lyu, S.~C.-K. Chau, N.~Wang, and Y.~Zheng, ``Cloud-based privacy-preserving collaborative consumption for sharing economy,'' \emph{IEEE Trans. Cloud Computing}, vol.~10, no.~3, pp. 1647--1660, 2022.

\bibitem{CZ22energyplan}
S.~C.-K. Chau and Y.~Zhou, ``Blockchain-enabled decentralized privacy-preserving group purchasing for retail energy plans,'' in \emph{Proc. ACM Intl. Conf. on Future Energy Systems (e-Energy)}, 2022, pp. 172--187.

\bibitem{ZC21energyplan}
Y.~Zhou and S.~C.-K. Chau, ``Sharing economy meets energy markets: Group purchasing of energy plans in retail energy markets,'' in \emph{ACM Intl. Conf. on Systems for Energy-Efficient Built Environments (BuildSys)}, 2021.

\bibitem{WCZ21energyshare}
N.~Wang, S.~C.-K. Chau, and Y.~Zhou, ``Privacy-preserving energy storage sharing with blockchain,'' in \emph{Proc. ACM Intl. Conf. on Future Energy Systems (e-Energy)}, 2021, pp. 185--198.

\bibitem{WCZ22energyshare}
------, ``Privacy-preserving energy storage sharing with blockchain and secure multi-party computation,'' \emph{ACM SIGEnergy Energy Informatics Review}, vol.~1, no.~1, pp. 32--50, 2021.

\bibitem{ZCGL22tiot}
H.~Zhu, S.~C.-K. Chau, G.~Guarddin, and W.~Liang, ``Integrating {IoT}-sensing and crowdsensing with privacy: Privacy-preserving hybrid sensing for smart cities,'' \emph{ACM Trans. Internet-of-Things}, vol.~3, no.~4, Sep 2022.

\bibitem{fuchsbauer2018algebraic}
G.~Fuchsbauer, E.~Kiltz, and J.~Loss, ``The algebraic group model and its applications,'' in \emph{{Annual International Cryptology Conference}}, 2018, pp. 33--62.

\bibitem{Fiat_Shamir_1987}
A.~Fiat and A.~Shamir, ``How to prove yourself: Practical solutions to identification and signature problems,'' in \emph{Advances in Cryptology --- CRYPTO' 86}, A.~M. Odlyzko, Ed.\hskip 1em plus 0.5em minus 0.4em\relax Berlin, Heidelberg: Springer Berlin Heidelberg, 1987, pp. 186--194.

\end{thebibliography}

\section*{Appendix}

\subsection{Restricted KZG Polynomial Commitment}\label{sec:rkzg}

We present a restricted version of KZG polynomial commitment scheme (denoted by ${\tt rKZG}$) in Table~\ref{tab:rkzg} that precludes a committed polynomial with a non-zero constant term. Given a Laurent polynomial $f[X] = \sum_{i = - d}^{d} a_i X^i$, this restricted polynomial commitment scheme does not allow the input $a_0$, and hence, a committed polynomial must have a zero constant term. Note that our scheme simplifies the one in \cite{maller2019sonic}, which also considers the degree of $f[X]$ bounded by a known constant less than $d$. 

\begin{table}[b!]
\caption{Restricted KZG Polynomial Commitment Scheme (${\tt rKZG}$)} \label{tab:rkzg} \vspace{-5pt}
\begin{longfbox}[border-break-style=none,border-color=\#bbbbbb,background-color=\#eeeeee,width=\linewidth]
\begin{itemize}[leftmargin=*]

    \item $\textsf{\footnotesize Setup}_{\tt rKZG}$: We suppose that there is a trusted party, who takes the security parameter ${\lambda}$ and generates $\mathbb G_1, \mathbb G_2, \mathbb G_{\mathcal S}$ with bilinear pairing ${\tt e}$. Then, it selects ${\tt g} \in {\mathbb G_1}, {\tt h} \in {\mathbb G_2}, {\alpha}, x \xleftarrow{\$} {\mathbb F_p}\backslash\{0, 1\}$ at random uniformly. Next, set the structured reference string as: 
	$${\tt srs} \leftarrow \big(({\tt g}^{x^i})_{i = - d}^{d}, ({\tt g}^{\alpha x^i})_{i = - d, i\neq 0}^{d}, ({\tt h},{\tt h}^{\alpha},{\tt h}^{\alpha x})\big)$$
	Note that ${\tt g}^{\alpha}$ is explicitly removed from ${\tt srs}$. 

    \item $\textsf{\footnotesize Commit}_{\tt rKZG}$:  Given a Laurent polynomial $f[X] = \sum_{i = - d, i \ne 0}^{d} a_i X^i$, set the commitment as: 
	$$F \leftarrow {\tt g}^{\alpha \cdot f[x]} = \prod_{i=-d, i \ne 0}^{d}  ({\tt g}^{\alpha x^{i}})^{a_i}$$

    \item $\textsf{\footnotesize Open}_{\tt rKZG}$: To generate a proof $\pi$ to the evaluation $v=f[z]$ for commitment $F$, compute polynomial $q[X] = \frac{f[X]-f[z]}{X-z}$ by a polynomial factorization algorithm. Suppose $q[X] = \sum_{i = - d}^{d} b_i X^i$. Then, set the proof by: 
	$$\pi  \leftarrow  {\tt g}^{q[x]} = \prod^{d}_{i=-d}  ({\tt g}^{{x}^i})^{b_i}$$
	    
    \item $\textsf{\footnotesize Verify}_{\tt rKZG}$: To verify $({\tt srs}, F, z, v, \pi)$, the verifier checks the following pairing equation:
	\begin{equation}
	{\tt e}\langle \pi,{\tt h}^{\alpha x}\rangle \cdot {\tt e}\langle {\tt g}^{v}\pi^{-z},{\tt h}^{\alpha}\rangle \stackrel{?}{=} {\tt e}\langle F,{\tt h}\rangle \label{eqn:rKZGbChkOrg}
	\end{equation}
	That is, checking ${\tt e}\langle {\tt g}, {\tt h}\rangle^{\alpha x \cdot (q[x]) + \alpha(v - z\cdot q[x])} \overset{?}{=} {\tt e}\langle {\tt g}, {\tt h}\rangle^{\alpha \cdot f[x]}$

\end{itemize}
\end{longfbox}
\end{table}

\subsection{Restricted Polynomial Commitment with Batch Verification}\label{sec:batchkzg}

We introduce a new restricted polynomial commitment scheme (denoted by ${\tt rKZGb}$), which verifies the openings of multiple polynomials at different evaluation points together. Thus, we can reduce the verification overhead in Sonic Protocol by batch verification. Our scheme extends the ideas in \cite{boneh21kzg} to the restricted KZG polynomial commitment scheme ${\tt rKZG}$. 

There are $K$ polynomials $\{ f_i[X] \}_{i=1}^K$. The prover commits to these polynomials as $F_i = \textsf{\footnotesize Commit}_{\tt rKZGb}(f_i[X])$. We omit the ${\tt srs}$ for the sake of brevity. There are a set of evaluation points ${\mathcal S} = \{z_1, , ..., , z_n \}$. Only the subset of evaluation points ${\mathcal S}_i \subset {\mathcal S}$ will be evaluated on the $i$-th polynomial $f_i[X]$. The prover aims to open $\{ (f_i[z])_{z \in {\mathcal S}_i} \}_{i=1}^K$ in a batch to the verifier, together with one single proof to prove all the openings are correct with respect to the commitments $(F_i)_{i=1}^K$.

\begin{table}[b!]
\caption{Restricted KZG Polynomial Commitment Scheme\\ with Batch Verification (${\tt rKZGb}$)} \label{tab:batchkzg} \vspace{-5pt}
\begin{longfbox}[border-break-style=none,border-color=\#bbbbbb,background-color=\#eeeeee,width=\linewidth]
\begin{itemize}[leftmargin=*]

    \item $\textsf{\footnotesize Setup}_{\tt rKZGb}$: We suppose that there is a trusted party, who takes the security parameter ${\lambda}$ and generates $\mathbb G_1, \mathbb G_2, \mathbb G_{\mathcal S}$ with bilinear pairing ${\tt e}$. Then, it randomly selects $g \in {\mathbb G_1}, h \in {\mathbb G_2}, ({\alpha, x}) \xleftarrow{\$} {\mathbb F_p}\backslash\{0, 1\}$. Next, set the structured reference string as: 
	$${\tt srs} \leftarrow \big(({\tt g}^{x^i})_{i = - d}^{d}, ({\tt g}^{\alpha x^i})_{i = - d, i\neq 0}^{d}, ({\tt h},{\tt h}^{\alpha},{\tt h}^{\alpha x})\big)$$

    \item $\textsf{\footnotesize Commit}_{\tt rKZGb}$:  Given a Laurent polynomial $f[X] = \sum_{i = - d, i \ne 0}^{d} a_i X^i$, set polynomial $\gamma_i[X]$ such that $\gamma_i[z] = f_i[z]$ for all $z \in {\mathcal S}_i$ by Lagrange's interpolation, and set the commitment as: 
	$$F \leftarrow {\tt g}^{\alpha f[x]} = \prod_{i=-d, i \ne 0}^{d}  ({\tt g}^{\alpha x^{i}})^{a_i}$$

    \item $\textsf{\footnotesize BatchOpen}_{\tt rKZGb}$: Given a set of $K$ polynomials $\{ f_i[X] \}_{i=1}^K$, to generate a proof for the evaluation $\{ (f_i[z])_{z\in {\mathcal S}_i} \}_{i=1}^K$ on commitments $(F_i)_{i=1}^K$, the prover computes polynomials $p[X] = \frac{\hat{f}[X]}{Z_{{\mathcal S}}[X]}$ given $\beta$, and then $w_\mu[X] = \frac{\ell_\mu[X]}{(X-\mu)}$ given $\mu$. Then, set the proof $(\pi_1, \pi_2)$ by: 
	$$\pi_1  \leftarrow {\tt g}^{p[x]}, \ \pi_2  \leftarrow  {\tt g}^{w_\mu[x]}$$
	    
    \item $\textsf{\footnotesize BatchVerify}_{\tt rKZGb}$: To verify $\big({\tt srs}, (F_i)_{i=1}^K, ({\mathcal S}_i)_{i=1}^K, \{ \gamma_i[X] \}_{i=1}^K, (\pi_1,\pi_2)\big)$, 
	\begin{enumerate}
	\item The verifier generates a random challenge $\beta \xleftarrow{\$} {\mathbb F_p}$.
	\item The verifier receives $\pi_1$ from the prover.
	\item The verifier generates a random challenge $\mu \xleftarrow{\$} {\mathbb F_p}$. 
	\item The verifier receives $\pi_2$ from the prover. 
	\item The verifier checks the following pairing equation:
	\begin{equation}
	{\tt e}\langle  \pi_2,\ {\tt h}^{\alpha x}\rangle \overset{?}{=} {\tt e}\Big\langle \Theta[\mu],\ h\Big\rangle \cdot {\tt e}\Big\langle \Phi'[\mu],\ {\tt h}^{\alpha}\Big\rangle \label{eqn:rKZGbChk}
	\end{equation}	
    where $\Psi_i[\mu] \triangleq \beta^{i-1}\cdot Z_{{\mathcal S}\backslash {\mathcal S}_i}[\mu]$,	
	\[ 
	\Theta[\mu] \triangleq \prod_{i=1}^K F_i^{\Psi_i[\mu]}, \ \ 
	\Phi'[\mu] \triangleq \frac{\pi_2^{\mu}}{\pi_1^{Z_{\mathcal S}[\mu]}} \prod_{i=1}^K {\tt g}^{-\gamma_i[\mu]\cdot\Psi_i[\mu]} 
	\]
	\end{enumerate}
\end{itemize}
\end{longfbox}
\end{table}

Define polynomial $\gamma_i[X]$, such that $\gamma_i[z] = f_i[z]$ for all $z \in {\mathcal S}_i$, which can be constructed by Lagrange's interpolation:
\[
\gamma_i[X] \triangleq \sum_{z \in {\mathcal S}_i} f_i[z] \cdot\Big(\frac{\prod_{z' \in {\mathcal S}_i\backslash\{z\}} (X-z')}{\prod_{z' \in {\mathcal S}_i\backslash\{z\}} (z-z')}\Big)
\]

Given a random challenge $\beta \xleftarrow{\$} {\mathbb F_p}$ from the verifier, define $Z_{S}[X] \triangleq \prod_{z \in S} (X-z)$ and 
\[
\hat{f}[X]  \triangleq \sum_{i=1}^K \beta^{i-1} \cdot Z_{{\mathcal S}\backslash {\mathcal S}_i}[X]\cdot(f_i[X] - \gamma_i[X])
\]
Note that 
\[
\frac{Z_{{\mathcal S}\backslash {\mathcal S}_i}[X]\cdot(f_i[X] - \gamma_i[X])}{Z_{{\mathcal S}}[X]} = \frac{f_i[X] - \gamma_i[X]}{Z_{{\mathcal S}_i}[X]} =  \frac{f_i[X] - \gamma_i[X]}{\prod_{z \in {\mathcal S}_i} (X-z)}
\]

Since $z \in {\mathcal S}_i$ are the roots for $f_i[X] - \gamma_i[X]$, and hence, $f_i[X] - \gamma_i[X]$ is divisible by $Z_{{\mathcal S}_i}[X]$. Therefore, $\hat{f}[X]$  is divisible by $Z_{{\mathcal S}}[X]$. Let $p[X] \triangleq \frac{\hat{f}[X]}{Z_{{\mathcal S}}[X]}$.

Given a random challenge $\mu \xleftarrow{\$} {\mathbb F_p}$ from the verifier, define:
\begin{align}
\hat{f}_\mu[X] & \triangleq \sum_{i=1}^K \beta^{i-1} \cdot Z_{{\mathcal S}\backslash {\mathcal S}_i}[\mu]\cdot(f_i[X] - \gamma_i[\mu]) \notag\\
\ell_\mu[X] & \triangleq \hat{f}_\mu[X] - \hat{f}[X] \notag
\end{align}
It is evident to see that $\ell_\mu[X]$ is divisible by $(X-\mu)$, since $\hat{f}_\mu[\mu] - \hat{f}[\mu] = 0$. Let $w_\mu[X] \triangleq \frac{\ell_\mu[X]}{(X-\mu)}$.

Next, we derive the following equation:
\begin{align}
       &{\tt g}^{\alpha \cdot w_\mu[x] \cdot (x-\mu)} = {\tt g}^{\alpha\cdot\ell_\mu[x]} = (\frac{{\tt g}^{\hat{f}_\mu[x]}}{{\tt g}^{\hat{f}[x]}})^{\alpha} \label{eqn:batchcheck1} \\ 
       =&\frac{1}{{\tt g}^{\alpha \cdot p[x]\cdot Z_{\mathcal S}[x]}}\prod_{i=1}^K (\frac{F_i }{{\tt g}^{\alpha \cdot\gamma_i[\mu]}})^{\beta^{i-1} \cdot Z_{{\mathcal S}\backslash {\mathcal S}_i}[\mu]} \\
       =&\frac{{\tt g}^{-\sum_{i=1}^K \gamma_i[\mu] \cdot\beta^{i-1}\cdot Z_{{\mathcal S}\backslash {\mathcal S}_i}[\mu]}}{{\tt g}^{\alpha \cdot p[x]\cdot Z_{\mathcal S}[x]}} \prod_{i=1}^K F_i^{\beta^{i-1} \cdot Z_{{\mathcal S}\backslash {\mathcal S}_i}[\mu]}  \label{eqn:batchcheck3}
\end{align}

Therefore, by Eqns.~(\ref{eqn:batchcheck1})-(\ref{eqn:batchcheck3}), we can validate the openings of $\{ (f_i[z])_{z \in {\mathcal S}_i} \}_{i=1}^K$ by the following pairing equation:
\[
{\tt e}\langle  {\tt g}^{w_\mu[x]},\ {\tt h}^{\alpha x}\rangle \overset{?}{=} {\tt e}\Big\langle \Theta[\mu],\ {\tt h}\Big\rangle \cdot {\tt e}\Big\langle \Phi[x,\mu],\ {\tt h}^{\alpha}\Big\rangle
\]
where $\Psi_i[\mu] \triangleq \beta^{i-1}\cdot Z_{{\mathcal S}\backslash {\mathcal S}_i}[\mu]$,	
\[ 
\Theta[\mu] \triangleq \prod_{i=1}^K F_i^{\Psi_i[\mu]}, \ \ 
\Phi[x,\mu] \triangleq \frac{{\tt g}^{\mu \cdot w_\mu[x]}}{{\tt g}^{p[x]\cdot Z_{\mathcal S}[\mu]}} \prod_{i=1}^K {\tt g}^{-\gamma_i[\mu]\cdot\Psi_i[\mu]} 
\]

We describe the restricted KZG polynomial commitment scheme with batch verification (${\tt rKZGb}$) in Table~\ref{tab:batchkzg}. Note that $\textsf{\footnotesize BatchVerify}_{\tt rKZGb}$ is an interactive process, which can be converted to a non-interactive one by Fiat-Shamir heuristic.

\section*{Appendix}

\subsection{Restricted KZG Polynomial Commitment}\label{sec:rkzg}

We present a restricted version of KZG polynomial commitment scheme (denoted by ${\tt rKZG}$) in Table~\ref{tab:rkzg} that precludes a committed polynomial with a non-zero constant term. Given a Laurent polynomial $f[X] = \sum_{i = - d}^{d} a_i X^i$, this restricted polynomial commitment scheme does not allow the input $a_0$, and hence, a committed polynomial must have a zero constant term. Note that our scheme simplifies the one in \cite{maller2019sonic}, which also considers the degree of $f[X]$ bounded by a known constant less than $d$. 

\begin{table}[b!]
\caption{Restricted KZG Polynomial Commitment Scheme (${\tt rKZG}$)} \label{tab:rkzg} \vspace{-5pt}
\begin{longfbox}[border-break-style=none,border-color=\#bbbbbb,background-color=\#eeeeee,width=\linewidth]
\begin{itemize}[leftmargin=*]

    \item $\textsf{\footnotesize Setup}_{\tt rKZG}$: We suppose that there is a trusted party, who takes the security parameter ${\lambda}$ and generates $\mathbb G_1, \mathbb G_2, \mathbb G_{\mathcal S}$ with bilinear pairing ${\tt e}$. Then, it selects ${\tt g} \in {\mathbb G_1}, {\tt h} \in {\mathbb G_2}, {\alpha}, x \xleftarrow{\$} {\mathbb F_p}\backslash\{0, 1\}$ at random uniformly. Next, set the structured reference string as: 
	$${\tt srs} \leftarrow \big(({\tt g}^{x^i})_{i = - d}^{d}, ({\tt g}^{\alpha x^i})_{i = - d, i\neq 0}^{d}, ({\tt h},{\tt h}^{\alpha},{\tt h}^{\alpha x})\big)$$
	Note that ${\tt g}^{\alpha}$ is explicitly removed from ${\tt srs}$. 

    \item $\textsf{\footnotesize Commit}_{\tt rKZG}$:  Given a Laurent polynomial $f[X] = \sum_{i = - d, i \ne 0}^{d} a_i X^i$, set the commitment as: 
	$$F \leftarrow {\tt g}^{\alpha \cdot f[x]} = \prod_{i=-d, i \ne 0}^{d}  ({\tt g}^{\alpha x^{i}})^{a_i}$$

    \item $\textsf{\footnotesize Open}_{\tt rKZG}$: To generate a proof $\pi$ to the evaluation $v=f[z]$ for commitment $F$, compute polynomial $q[X] = \frac{f[X]-f[z]}{X-z}$ by a polynomial factorization algorithm. Suppose $q[X] = \sum_{i = - d}^{d} b_i X^i$. Then, set the proof by: 
	$$\pi  \leftarrow  {\tt g}^{q[x]} = \prod^{d}_{i=-d}  ({\tt g}^{{x}^i})^{b_i}$$
	    
    \item $\textsf{\footnotesize Verify}_{\tt rKZG}$: To verify $({\tt srs}, F, z, v, \pi)$, the verifier checks the following pairing equation:
	\begin{equation}
	{\tt e}\langle \pi,{\tt h}^{\alpha x}\rangle \cdot {\tt e}\langle {\tt g}^{v}\pi^{-z},{\tt h}^{\alpha}\rangle \stackrel{?}{=} {\tt e}\langle F,{\tt h}\rangle \label{eqn:rKZGbChkOrg}
	\end{equation}
	That is, checking ${\tt e}\langle {\tt g}, {\tt h}\rangle^{\alpha x \cdot (q[x]) + \alpha(v - z\cdot q[x])} \overset{?}{=} {\tt e}\langle {\tt g}, {\tt h}\rangle^{\alpha \cdot f[x]}$

\end{itemize}
\end{longfbox}
\end{table}

\subsection{Restricted Polynomial Commitment with Batch Verification}\label{sec:batchkzg}

We introduce a new restricted polynomial commitment scheme (denoted by ${\tt rKZGb}$), which verifies the openings of multiple polynomials at different evaluation points together. Thus, we can reduce the verification overhead in Sonic Protocol by batch verification. Our scheme extends the ideas in \cite{boneh21kzg} to the restricted KZG polynomial commitment scheme ${\tt rKZG}$. 

There are $K$ polynomials $\{ f_i[X] \}_{i=1}^K$. The prover commits to these polynomials as $F_i = \textsf{\footnotesize Commit}_{\tt rKZGb}(f_i[X])$. We omit the ${\tt srs}$ for the sake of brevity. There are a set of evaluation points ${\mathcal S} = \{z_1, , ..., , z_n \}$. Only the subset of evaluation points ${\mathcal S}_i \subset {\mathcal S}$ will be evaluated on the $i$-th polynomial $f_i[X]$. The prover aims to open $\{ (f_i[z])_{z \in {\mathcal S}_i} \}_{i=1}^K$ in a batch to the verifier, together with one single proof to prove all the openings are correct with respect to the commitments $(F_i)_{i=1}^K$.

\begin{table}[b!]
\caption{Restricted KZG Polynomial Commitment Scheme\\ with Batch Verification (${\tt rKZGb}$)} \label{tab:batchkzg} \vspace{-5pt}
\begin{longfbox}[border-break-style=none,border-color=\#bbbbbb,background-color=\#eeeeee,width=\linewidth]
\begin{itemize}[leftmargin=*]

    \item $\textsf{\footnotesize Setup}_{\tt rKZGb}$: We suppose that there is a trusted party, who takes the security parameter ${\lambda}$ and generates $\mathbb G_1, \mathbb G_2, \mathbb G_{\mathcal S}$ with bilinear pairing ${\tt e}$. Then, it randomly selects $g \in {\mathbb G_1}, h \in {\mathbb G_2}, ({\alpha, x}) \xleftarrow{\$} {\mathbb F_p}\backslash\{0, 1\}$. Next, set the structured reference string as: 
	$${\tt srs} \leftarrow \big(({\tt g}^{x^i})_{i = - d}^{d}, ({\tt g}^{\alpha x^i})_{i = - d, i\neq 0}^{d}, ({\tt h},{\tt h}^{\alpha},{\tt h}^{\alpha x})\big)$$

    \item $\textsf{\footnotesize Commit}_{\tt rKZGb}$:  Given a Laurent polynomial $f[X] = \sum_{i = - d, i \ne 0}^{d} a_i X^i$, set polynomial $\gamma_i[X]$ such that $\gamma_i[z] = f_i[z]$ for all $z \in {\mathcal S}_i$ by Lagrange's interpolation, and set the commitment as: 
	$$F \leftarrow {\tt g}^{\alpha f[x]} = \prod_{i=-d, i \ne 0}^{d}  ({\tt g}^{\alpha x^{i}})^{a_i}$$

    \item $\textsf{\footnotesize BatchOpen}_{\tt rKZGb}$: Given a set of $K$ polynomials $\{ f_i[X] \}_{i=1}^K$, to generate a proof for the evaluation $\{ (f_i[z])_{z\in {\mathcal S}_i} \}_{i=1}^K$ on commitments $(F_i)_{i=1}^K$, the prover computes polynomials $p[X] = \frac{\hat{f}[X]}{Z_{{\mathcal S}}[X]}$ given $\beta$, and then $w_\mu[X] = \frac{\ell_\mu[X]}{(X-\mu)}$ given $\mu$. Then, set the proof $(\pi_1, \pi_2)$ by: 
	$$\pi_1  \leftarrow {\tt g}^{p[x]}, \ \pi_2  \leftarrow  {\tt g}^{w_\mu[x]}$$
	    
    \item $\textsf{\footnotesize BatchVerify}_{\tt rKZGb}$: To verify $\big({\tt srs}, (F_i)_{i=1}^K, ({\mathcal S}_i)_{i=1}^K, \{ \gamma_i[X] \}_{i=1}^K, (\pi_1,\pi_2)\big)$, 
	\begin{enumerate}
	\item The verifier generates a random challenge $\beta \xleftarrow{\$} {\mathbb F_p}$.
	\item The verifier receives $\pi_1$ from the prover.
	\item The verifier generates a random challenge $\mu \xleftarrow{\$} {\mathbb F_p}$. 
	\item The verifier receives $\pi_2$ from the prover. 
	\item The verifier checks the following pairing equation:
	\begin{equation}
	{\tt e}\langle  \pi_2,\ {\tt h}^{\alpha x}\rangle \overset{?}{=} {\tt e}\Big\langle \Theta[\mu],\ h\Big\rangle \cdot {\tt e}\Big\langle \Phi'[\mu],\ {\tt h}^{\alpha}\Big\rangle \label{eqn:rKZGbChk}
	\end{equation}	
    where $\Psi_i[\mu] \triangleq \beta^{i-1}\cdot Z_{{\mathcal S}\backslash {\mathcal S}_i}[\mu]$,	
	\[ 
	\Theta[\mu] \triangleq \prod_{i=1}^K F_i^{\Psi_i[\mu]}, \ \ 
	\Phi'[\mu] \triangleq \frac{\pi_2^{\mu}}{\pi_1^{Z_{\mathcal S}[\mu]}} \prod_{i=1}^K {\tt g}^{-\gamma_i[\mu]\cdot\Psi_i[\mu]} 
	\]
	\end{enumerate}
\end{itemize}
\end{longfbox}
\end{table}

Define polynomial $\gamma_i[X]$, such that $\gamma_i[z] = f_i[z]$ for all $z \in {\mathcal S}_i$, which can be constructed by Lagrange's interpolation:
\[
\gamma_i[X] \triangleq \sum_{z \in {\mathcal S}_i} f_i[z] \cdot\Big(\frac{\prod_{z' \in {\mathcal S}_i\backslash\{z\}} (X-z')}{\prod_{z' \in {\mathcal S}_i\backslash\{z\}} (z-z')}\Big)
\]

Given a random challenge $\beta \xleftarrow{\$} {\mathbb F_p}$ from the verifier, define $Z_{S}[X] \triangleq \prod_{z \in S} (X-z)$ and 
\[
\hat{f}[X]  \triangleq \sum_{i=1}^K \beta^{i-1} \cdot Z_{{\mathcal S}\backslash {\mathcal S}_i}[X]\cdot(f_i[X] - \gamma_i[X])
\]
Note that 
\[
\frac{Z_{{\mathcal S}\backslash {\mathcal S}_i}[X]\cdot(f_i[X] - \gamma_i[X])}{Z_{{\mathcal S}}[X]} = \frac{f_i[X] - \gamma_i[X]}{Z_{{\mathcal S}_i}[X]} =  \frac{f_i[X] - \gamma_i[X]}{\prod_{z \in {\mathcal S}_i} (X-z)}
\]

Since $z \in {\mathcal S}_i$ are the roots for $f_i[X] - \gamma_i[X]$, and hence, $f_i[X] - \gamma_i[X]$ is divisible by $Z_{{\mathcal S}_i}[X]$. Therefore, $\hat{f}[X]$  is divisible by $Z_{{\mathcal S}}[X]$. Let $p[X] \triangleq \frac{\hat{f}[X]}{Z_{{\mathcal S}}[X]}$.

Given a random challenge $\mu \xleftarrow{\$} {\mathbb F_p}$ from the verifier, define:
\begin{align}
\hat{f}_\mu[X] & \triangleq \sum_{i=1}^K \beta^{i-1} \cdot Z_{{\mathcal S}\backslash {\mathcal S}_i}[\mu]\cdot(f_i[X] - \gamma_i[\mu]) \notag\\
\ell_\mu[X] & \triangleq \hat{f}_\mu[X] - \hat{f}[X] \notag
\end{align}
It is evident to see that $\ell_\mu[X]$ is divisible by $(X-\mu)$, since $\hat{f}_\mu[\mu] - \hat{f}[\mu] = 0$. Let $w_\mu[X] \triangleq \frac{\ell_\mu[X]}{(X-\mu)}$.

Next, we derive the following equation:
\begin{align}
       &{\tt g}^{\alpha \cdot w_\mu[x] \cdot (x-\mu)} = {\tt g}^{\alpha\cdot\ell_\mu[x]} = (\frac{{\tt g}^{\hat{f}_\mu[x]}}{{\tt g}^{\hat{f}[x]}})^{\alpha} \label{eqn:batchcheck1} \\ 
       =&\frac{1}{{\tt g}^{\alpha \cdot p[x]\cdot Z_{\mathcal S}[x]}}\prod_{i=1}^K (\frac{F_i }{{\tt g}^{\alpha \cdot\gamma_i[\mu]}})^{\beta^{i-1} \cdot Z_{{\mathcal S}\backslash {\mathcal S}_i}[\mu]} \\
       =&\frac{{\tt g}^{-\sum_{i=1}^K \gamma_i[\mu] \cdot\beta^{i-1}\cdot Z_{{\mathcal S}\backslash {\mathcal S}_i}[\mu]}}{{\tt g}^{\alpha \cdot p[x]\cdot Z_{\mathcal S}[x]}} \prod_{i=1}^K F_i^{\beta^{i-1} \cdot Z_{{\mathcal S}\backslash {\mathcal S}_i}[\mu]}  \label{eqn:batchcheck3}
\end{align}

Therefore, by Eqns.~(\ref{eqn:batchcheck1})-(\ref{eqn:batchcheck3}), we can validate the openings of $\{ (f_i[z])_{z \in {\mathcal S}_i} \}_{i=1}^K$ by the following pairing equation:
\[
{\tt e}\langle  {\tt g}^{w_\mu[x]},\ {\tt h}^{\alpha x}\rangle \overset{?}{=} {\tt e}\Big\langle \Theta[\mu],\ {\tt h}\Big\rangle \cdot {\tt e}\Big\langle \Phi[x,\mu],\ {\tt h}^{\alpha}\Big\rangle
\]
where $\Psi_i[\mu] \triangleq \beta^{i-1}\cdot Z_{{\mathcal S}\backslash {\mathcal S}_i}[\mu]$,	
\[ 
\Theta[\mu] \triangleq \prod_{i=1}^K F_i^{\Psi_i[\mu]}, \ \ 
\Phi[x,\mu] \triangleq \frac{{\tt g}^{\mu \cdot w_\mu[x]}}{{\tt g}^{p[x]\cdot Z_{\mathcal S}[\mu]}} \prod_{i=1}^K {\tt g}^{-\gamma_i[\mu]\cdot\Psi_i[\mu]} 
\]

We describe the restricted KZG polynomial commitment scheme with batch verification (${\tt rKZGb}$) in Table~\ref{tab:batchkzg}. Note that $\textsf{\footnotesize BatchVerify}_{\tt rKZGb}$ is an interactive process, which can be converted to a non-interactive one by Fiat-Shamir heuristic.  
\\

\clearpage

\subsection{Proof of Knowledge Soundness for ${\tt rKZGb}$}

We next prove some desirable properties of our polynomial commitment scheme ${\tt rKZGb}$ by the following theorem.

\begin{theorem}
\label{thm:enhanced-kzg-requirement} \label{thm:rKZGb}
Polynomial commitment scheme ${\tt rKZGb}$ satisfies correctness, knowledge soundness and computational hiding.
\end{theorem}

\textit{Proof}: 
We only prove knowledge soundness, as correctness follows from Eqns.~(\ref{eqn:batchcheck1})-(\ref{eqn:batchcheck3}) and computational hiding follows from computational Diffie-Hellman assumption.
Informally, knowledge soundness means for every efficient adversary $\mathcal A$ who can pass the verification successfully, there exists an efficient extractor ${\mathcal E}_{\mathcal A}$ who can extract the committed polynomial with high probability given the access to ${\mathcal A}$'s internal states.

Next, we define the adversary $\mathcal A$ properly. Traditionally, $\mathcal A$ is defined by Generic Group Model (GGM), who treats ${\tt srs}$ as a blackbox. Here we consider a more powerful adversary, who can utilize efficient group representation to generate a polynomial commitment from ${\tt srs}$ for a successful verification. We define the adversary by Algebraic Group Model (AGM). We only provide a simple argument of AGM. The details of AGM can be found in to \cite{fuchsbauer2018algebraic}.

We adopt the following definition of algebraic adversary and $q$-DLOG assumption from \cite{boneh21kzg, fuchsbauer2018algebraic}.

\begin{definition}[\textit{Algebraic Adversary}]
An algebraic adversary $\mathcal A$ is a polynomial-time algorithm that when $\mathcal A$ is asked to output a commitment $F \in \mathbb G_1$ or $\mathbb G_2$, it also outputs a vector of linear combination of elements in ${\tt srs}$, such that $F = \prod_{t=1}^{N} {\tt srs}_{t}^{a_{t}}$, where $({\tt srs}_{t})_{t=1}^{N}$ are the elements in ${\tt srs}$.
\end{definition}

\begin{definition}[\textit{q-DLOG}]
Given ${\tt srs} \leftarrow \textsf{\footnotesize\em Setup}_{\tt rKZGb}(\lambda, {\alpha, x})$, where $({\alpha, x}) \xleftarrow{\$} {\mathbb F_p}\backslash\{0, 1\}$, the probability that any algebraic adversary $\mathcal A$ can output $({\alpha, x})$ efficiently from ${\tt srs}$ is negligible $\mathbf{\epsilon}(\lambda)$. 
\end{definition}

Formally, we define knowledge soundness in AGM for polynomial commitment scheme ${\tt rKZGb}$ by a game with ${\tt rKZGb}$ that the probability of $\mathcal A$'s winning (by providing false openings to pass $\textsf{\footnotesize BatchVerify}_{\tt rKZGb}$) is negligible.

\begin{definition}[\textit{Knowledge Soundness in AGM}] Assuming $q$-DLOG, polynomial commitment scheme ${\tt rKZGb}$ satisfies knowledge soundness, if there exists an efficient extractor $\mathcal{E}_\mathcal{A}$ for any algebraic adversary $\mathcal{A}$, when the probability of $\mathcal{A}$'s winning in the following game is negligible:
\begin{enumerate}
  \item Given ${\tt srs} \leftarrow \textsf{\footnotesize\em Setup}_{\tt rKZGb}(\lambda, \alpha, x)$, $\mathcal{A}$ outputs a set of commitments $(F_i)^K_{i=1}$, such that each $F_i = \prod_{t=1}^{N} {\tt srs}_{t}^{a_{t,i}}$.

  \item Extractor $\mathcal{E}_\mathcal{A}$, given access to $\mathcal{A}$'s internal states, extract polynomials $(f_i[X])^K_{i=1}$. 

  \item $\mathcal A$ provides $\big(({\mathcal S}_i)_{i=1}^K, \{ \gamma_i[X] \}_{i=1}^K\big)$.
  
  \item The verifier generates a random challenge $\beta \xleftarrow{\$} {\mathbb F_p}$.
  
  \item $\mathcal A$ provides $\pi_1$.
  
  \item The verifier generates a random challenge $\mu \xleftarrow{\$} {\mathbb F_p}$.
  
  \item $\mathcal A$ provides $\pi_2$.
 
  \item $\mathcal{A}$ wins if proof $(\pi_1, \pi_2)$ passes $\textsf{\footnotesize\em BatchVerify}_{\tt rKZGb}$, but there exists ${j} \in \{1, ..., K\}, z \in {\mathcal S}_j$, such that $f_{j}[z] \neq \gamma_{j}[z]$.
\end{enumerate}
\end{definition}

We follow a similar argument in \cite{gabizon2019plonk}. Let us assume that such a winning $\mathcal A$ exists. Note that $f_{j}[z] \neq \gamma_{j}[z]$ is equivalent to $\left(f_{j}[X]-\gamma_{j}[X]\right)$ being indivisible by $Z_{{\mathcal S}_{j}}[X]$.

Since $\mathcal{E}_{\mathcal A}$ has access to $\mathcal A$'s internal states, when $\mathcal A$ ouputs $F_i = \prod_{i=-d}^{d} ({\tt g}^{\alpha x^{i}})^{a_i}$, then $\mathcal{E}_{\mathcal A}$ extracts $f_i[X] = \sum^{d}_{-d} a_i X^i$. 

In $\textsf{\footnotesize BatchVerify}_{\tt rKZGb}$, the verifier generates $\beta \xleftarrow{\$} {\mathbb F_p}$. Recall 
$$
\hat{f}[X]\triangleq \sum_{i=1}^K \beta^{i-1} \cdot Z_{{\mathcal S} \backslash {\mathcal S}_{i}} \cdot\left(f_{i}[X]-\gamma_{i}[X]\right)
$$
Note that $f[X]$ being divisible by $Z_{{\mathcal S}_{i}}[X]$ is equivalent to $Z_{{\mathcal S} \backslash {\mathcal S}_{j}}[X] \cdot f[X]$ being divisible by $Z_{\mathcal S}[X]$, and $(f_{j}[X]-r_{j}[X])$ being indivisible by $Z_{{\mathcal S}_{j}}[X]$ is equivalent to $Z_{{\mathcal S} \backslash {\mathcal S}_{j}} \cdot (f_{j}[X]-\gamma_{j}[X])$ being indivisible by $Z_{{\mathcal S}}[X]$. 

We adopt the following lemma adapted from \cite{gabizon2019plonk}.

\begin{lemma}[\cite{gabizon2019plonk} Proved Claim 4.6]
Given polynomial $F[X] = \sum_{i=1}^K \beta^{i-1} \cdot f_i[X]$, where $f_i[X]$ is a polynomial over a finite field $\mathbb F$ with a degree bounded in $[-d, d]$, and $Z[X]$ that decomposes to distinct linear factors over $\mathbb F$. Suppose for some $j \in \{1, ..., K\}$, $f_j[X]$ is indivisible by $Z[X]$, then with probability at least $1-\frac{K}{|\mathbb F|}$, $F[X]$ is also indivisible by $Z[X]$.
\label{lem:indivisible}
\end{lemma}

By Lemma~\ref{lem:indivisible}, we obtain that $\hat{f}[X]$ is indivisible by $Z_{{\mathcal S}}[X]$ with probability at least $1-\frac{K}{|\mathbb F_p|}$. 
If $\hat{f}[X]$ is divisible by $Z_{{\mathcal S}}[X]$, then $\mathcal A$ wins, but with a negligible probability $\frac{K}{|\mathbb F_p|}$, as $|\mathbb F_p|$ increases with $\lambda$. Suppose $\hat{f}[X]$ is indivisible by $Z_{{\mathcal S}}[X]$  - which we call Assumption ($\ast$). Then $\mathcal A$ provides $\tilde{p}[X]$ in $\pi_1$, but it is certain that $\hat{f}[X] \ne \tilde{p}[X] \cdot Z_{{\mathcal S}}[X]$.

Next, the verifier generates $\mu \xleftarrow{\$} {\mathbb F_p}$, and $\mathcal A$ provides $\hat{w}_\mu[X]$ in $\pi_2$. If $\mathcal A$ wins, then the proof $(\pi_1, \pi_2) = ({\tt g}^{\tilde{p}[x]}, {\tt g}^{\hat{w}_\mu[x]})$ must pass the pairing Eqn.~(\ref{eqn:rKZGbChk}) in $\textsf{\footnotesize BatchVerify}_{\tt rKZGb}$.

We define a ``\textit{real pairing check}'' with a group element $F$ that means checking $F$ by invoking some pairing equation with ${\tt e}\langle \cdot, \cdot\rangle$, whereas the corresponding ``\textit{ideal pairing check}'' means checking $F$ instead with the vector $(a_t)_{t=1}^{N}$ where $F = \prod_{t=1}^{N} {\tt srs}_{t}^{a_{t}}$ by some linear equation.
We adopt the following Lemma from \cite{gabizon2019plonk}. 

\begin{lemma}[\cite{gabizon2019plonk} Lemma 2.2]
\label{lemma:realideal}
Assuming $q$-DLOG, for any algebraic adversary $\mathcal A$, the probability of passing a real pairing check is larger than the probability of passing the corresponding ideal pairing check by at most negligible $\mathbf{\epsilon}(\lambda)$.
\end{lemma}

By Lemma~\ref {lemma:realideal}, we can replace Eqn.~(\ref{eqn:rKZGbChk}) by the following ideal pairing check with a negligible difference in probability:
\begin{align}
& & \alpha \cdot \tilde{w}_\mu[x] \cdot (x-\mu) &\overset{?}{=}  \alpha \cdot \hat{f}[\mu] -\alpha \cdot \tilde{p}[x]\cdot Z_{\mathcal S}[x] \\
& \Rightarrow & \tilde{w}_\mu[x] \cdot (x-\mu) & \overset{?}{=}   \hat{f}[\mu] - \tilde{p}[x]\cdot Z_{\mathcal S}[x] \label{eqn:satisfyeq}
\end{align}

If $\mathcal A$ passes the ideal pairing check for an unknown $x$, then by Schwartz-Zippel Lemma \cite{crytobk}, $(\hat{f}[\mu] - \tilde{p}[x]\cdot Z_{\mathcal S}[x])$ is indivisible by $(x-\mu)$ with a small probability $O(\frac{1}{|{\mathbb F}_p|})$ (as Eqn~(\ref{eqn:satisfyeq}) is only satisfied for a very small set of points $x$ in ${\mathbb F}_p$, if $(\hat{f}[\mu] - \tilde{p}[x]\cdot Z_{\mathcal S}[x])$ is indivisible by $(x-\mu)$). If $(\hat{f}[\mu] - \tilde{p}[x]\cdot Z_{\mathcal S}[x])$ is divisible by $(x-\mu)$, then $\hat{f}[\mu] - \tilde{p}[\mu]\cdot Z_{\mathcal S}[\mu] = 0$, which contradicts the assumptions that $\hat{f}[X] \ne \tilde{p}[X] \cdot Z_{{\mathcal S}}[X]$ and $\hat{f}[X]$ are indivisible by $Z_{{\mathcal S}}[X]$ (Assumption ($\ast$)).

Therefore, $\mathcal A$ only wins with a negligible probability. This completes the proof for the knowledge soundness of polynomial commitment scheme ${\tt rKZGb}$ in AGM.
\hfill{$\square$}

\begin{table}[t!]
\caption{Enhanced Interactive zk-SNARK Protocol with Heterogeneous Data Sources and Batch Verification (${\tt S}_{\tt EV}$)} \label{tab:final_protocol} \vspace{-10pt}
\begin{longfbox}[border-break-style=none,border-color=\#bbbbbb,background-color=\#eeeeee,breakable=true,,width=\linewidth]
\textbf{Public Input}: $\lambda, \hat{s}[X,Y], \hat{k}[Y], ({\tt pk}_j)_{j=1}^J$ \\
\textbf{Data Source $j \in \{1, , ..., , J\}$}: $d_j[X], {\tt sk}_j$ \\
\textbf{Prover's Input}: $r[X,Y]$ \\
\textbf{Interactive zk-SNARK Protocol}:
\begin{enumerate}
\item 
{\rm Setup}: \\
\mbox{\color{gray}\it// Only store necessary ${\tt srs}$ elements on chain} \\
\mbox{\color{gray}\it// or store SRS by an oracle and retrieve from it when needed} \\
${\tt srs}\leftarrow \textsf{\scriptsize Setup}_{\tt rKZGb}(\lambda), {\tt srs}_j\leftarrow \textsf{\scriptsize Setup}_{\tt rKZGb}(\lambda)$ 
\\

\item 
${\rm Data}_j \Rightarrow {\rm Prover}$:
\[
(D_j, \gamma_{D_j})\leftarrow{\textsf{\scriptsize Commit}_{\tt rKZGb}}({\tt srs}_j,d_j[X]),\  \sigma_j\leftarrow{\textsf{\scriptsize Sign}}({\tt sk}_j,D_j)
\]

\item
{\rm Verifier} $\Rightarrow$ {\rm Prover}: $y\xleftarrow{\$}\mathbb {\mathbb F}_p$, $\beta \xleftarrow{\$}\mathbb {\mathbb F}_p$ {\color{gray}\it \\
// (Fiat-Shamir): $y \leftarrow{\textsf{\scriptsize Hash}}(D_1| , ..., |D_J), \beta \leftarrow{\textsf{\scriptsize Hash}'}(D_1| , ..., |D_J)$} \\

\item 
{\rm Prover} $\Rightarrow$ {\rm Verifier}:
{\color{gray}//${\mathcal S}_X,\hat{k},\hat{s},\hat{s}_1,\hat{s}_2$ computing is outsourced to prover:}
\begin{align*}
(D_j, \gamma_{D_j}, \sigma_j)_{j=1}^J&\\
({\mathcal S}_Y,\gamma_{{\mathcal S}_y})&\leftarrow{\textsf{\scriptsize Commit}_{\tt rKZGb}}({\tt srs},\hat{s}[1,Y]) \\
(K,\gamma_K)&\leftarrow{\textsf{\scriptsize Commit}_{\tt rKZGb}}({\tt srs},\hat{k}[Y]) \\
(R,\gamma_R)&\leftarrow{\textsf{\scriptsize Commit}_{\tt rKZGb}}({\tt srs},r[X,1]) \\
({\mathcal S},\gamma_{\mathcal S})&\leftarrow{\textsf{\scriptsize Commit}_{\tt rKZGb}}({\tt srs},t[X,y]) \\
(\tilde{R},\gamma_{\tilde{R}})&\leftarrow{\textsf{\scriptsize Commit}_{\tt rKZGb}}({\tt srs},\tilde{r}[X,1]) \\
({\mathcal S}_X,\gamma_{s_x})&\leftarrow{\textsf{\scriptsize Commit}_{\tt rKZGb}}({\tt srs},\hat{s}[X,y])  
\end{align*}

\item
{\rm Verifier} $\Rightarrow$ {\rm Prover}: $ z\xleftarrow{\$} {\mathbb F}_p$ {\color{gray}\it \\
// (Fiat-Shamir): $z\leftarrow{\textsf{\scriptsize Hash}}(D_1| , ..., |D_J|{\mathcal S}_Y|K|R|{\mathcal S}|\tilde{R}|{\mathcal S}_X)$} \\

\item 
{\rm Prover} $\Rightarrow$ {\rm Verifier}:
\begin{align*}
(\pi_1, \pi_2) &\leftarrow {\textsf{\scriptsize BatchOpen}_{\tt rKZGb}}\Big(\tt srs, \\
&\{f_i[X]\}_{i=1}^K=\Big\{\{d_j[X]\}_{j=1}^J,\tilde{r} [X,1], r[X,1],t[X,y],\hat{k}[Y], \\& \qquad \qquad \qquad  \quad \hat{s}[X,y],\hat{s}[1,Y] \Big\}, \\
&\{\gamma_i[X]\}_{i=1}^K=\Big\{ \{\gamma_{D_j}\}_{j=1}^J, \gamma_{\tilde{R}}, \gamma_R, \gamma_{\mathcal S}, \gamma_K, \gamma_{s_x}, \gamma_{{\mathcal S}_y} \Big\} \Big)
\end{align*}
\begin{align*}
&r_1\leftarrow r[z,1],
t\leftarrow t[z,y],
\tilde{r}\leftarrow \tilde{r}[z,1], 
r_2\leftarrow r[zy,1], d_j\leftarrow d_j[z], \forall j\\
&\hat{s}\leftarrow \hat{s}[z,y],
k\leftarrow \hat{k}[y], \hat{s}_1\leftarrow \hat{s}[1,y],
\hat{s}_2 \leftarrow \hat{s}[1,y] \\ 
\end{align*}

\item
Verifier checks:
\[
\begin{array}{l@{}r}
 \bigwedge_{j=1}^J   {\textsf{\scriptsize VerifySign}}({\tt pk}_j,D_j,\sigma_j) \wedge \bigwedge_{j=1}^J  {\textsf{\scriptsize Verify}}({\tt srs}_j,D_j,z,d_j,\pi_{d_j}) \wedge \\
 \big(r_1 \stackrel{?}{=} \tilde{r} + \sum_{j=1}^J d_j z^{N + \sum_{j=1}^{J-1} m _j}\big)  \wedge\\
 \big(t \stackrel{?}{=} r_1(r_2+\hat{s})-k\big) \wedge (\hat{s}_1 \stackrel{?}{=} \hat{s}_2) \wedge\\
 {\textsf{\scriptsize BatchVerify}}_{\tt rKZGb}\Big({\tt srs},F_i^K=\big\{\{D_j\}_{j=1}^J ,\tilde{R}, R, {\mathcal S}, K, {\mathcal S}_X, {\mathcal S}_Y\big\}, \\
 \quad {\mathcal S}_i^K=\big\{\{z\},\{z\},\{z,zy\},\{z\}, \{y\},\{z, 1\},\{y\}\big\}, \\
 \quad \{\gamma_i[X]\}_{i=1}^K=[\bigwedge_{j=1}^J{\gamma_{D_j}}, \gamma_{\tilde{R}}, \gamma_R, \gamma_{\mathcal S}, \gamma_K, \gamma_{s_x}, \gamma_{{\mathcal S}_y}], \\
 \quad (\pi_1, \pi_2)\Big)
\end{array}
\]

\end{enumerate}    
\end{longfbox} \vspace{-20pt}
\end{table}

\begin{table*}[t]
    \centering
	 \caption{Settings of vectors $({\mathbf u_q, \mathbf  v_q, \mathbf  w_q}, k_q)$ in the constraint system for parametric bushfire insurance. $\mathbf I(i, j)$ represents a vector of length $(10n+(n+2)k)$ for which every entry are $0$ except $i_{th}, j_{th}$ entries which are $1$.} \vspace{-10pt}
	 {\scriptsize
    \begin{tabular}{@{}c@{}||@{}c@{}|@{}c@{}|@{}c@{}|@{}c@{}}
    \hline
        Computation&$k_q$&${\bf u}_q$&${\bf v}_q$&${\bf w}_q$ \\
        \hline
        $\mathbf r^-_i+\mathbf s^-_i, i\in \{1,...,n\}$&
        $0$ 
        & $\mathbf I(6n+(n+2)k+i, 7n+(n+2)k+i)$
        & $\mathbf{-I}(i)$
        & $\mathbf I$\\
        $\mathbf r^-_i-\mathbf s^-_i-\mathbf \theta^-_i, i\in \{1,...,n\}$&
        $0$ 
        & $\mathbf I(6n+(n+2)k+i) - I(7n+(n+2)k+i,3n+i)$
        & $\mathbf I$
        & $\mathbf{-I}(i)$\\
        $\mathbf r^+_i+\mathbf s^+_i, i\in \{1,...,n\}$&
        $0$ 
        & $\mathbf I(8n+(n+2)k+i, 9n+(n+2)k+i)$
        & $\mathbf{-I}(n+i)$
        & $\mathbf I$\\
        $\mathbf r^+_i-\mathbf s^+_i-\mathbf \theta^+_i, i\in \{1,...,n\}$&
        $0$ 
        & $\mathbf I(8n+(n+2)k+i) - I(9n+(n+2)k+i,4n+i)$
        & $\mathbf I$
        & $\mathbf{-I}(n+i)$\\
        
        $\theta_{\max} = \sum^n_{i=1} (\mathbf \theta^-_i)^2 + \sum^n_{i=1} (\mathbf \theta^+_i)^2 + \theta_{d}$ &
        $\theta_{\max}$ 
        & $\mathbf I(6n+nk+1, ..., 6n+nk+k)$
        & $\mathbf I$
        & $\mathbf I(3n+1, ..., 5n)$\\

        $\mathbf n_i^- - \mathbf n_i^+ - \kappa, i\in \{1,...,n\}$&
        $\kappa$ 
        & $\mathbf I(i)-I(n+i)$
        & $\mathbf{-I}(2n+i)$
        & $\mathbf I$\\
        $\sum^k_{j=1} e_{i,j}, i\in \{1,...,n\}$&
        $0$ 
        & $\mathbf I(6n+(i-1)k+1, ..., 6n+ik)$
        & $\mathbf I$
        & $\mathbf{-I}(2n+i)$\\
        $\sum^n_{i=1} \mathbf i_i-G=\epsilon$&
        $\epsilon$ 
        & $\mathbf I(5n+1, ..., 6n)- \mathbf I(6n+(n+1)k+1, ..., 6n+(n+2)k )$
        & $\mathbf I$
        & $\mathbf I$\\

        Checking  $\mathbf i_i$ in $\mathbf a, \mathbf b, i\in \{1,...,n\}$&
        $1$ 
        & $\mathbf I(5n+i)$
        & $\mathbf I(5n+i)$
        & $\mathbf I$\\
          
        Checking  $e_{i,j}$ in $\mathbf a, \mathbf b, i\in \{1,...,n\}, j\in \{1,...,k\}$&
        $2^{j-1}$ 
        & $\mathbf I(6n+(i-1)k+j)$
        & $\mathbf I(6n+(i-1)k+j)$
        & $\mathbf I$\\
       
        Checking  $e_{\theta_{d},j}$ in $\mathbf a, \mathbf b, j\in \{1,...,k\}$&
        $2^{j-1}$ 
        & $\mathbf I(6n+nk+j)$
        & $\mathbf I(6n+nk+j)$
        & $\mathbf I$\\
        Checking  $e_{G,j}$ in $\mathbf a, \mathbf b, j\in \{1,...,k\}$&
        $2^{j-1}$ 
        & $\mathbf I(6n+(n+1)k+j)$
        & $\mathbf I(6n+(n+1)k+j)$
        & $\mathbf I$\\
        
    \hline
    \end{tabular}
	}
    \label{tab:linear_const} \vspace{-10pt}
\end{table*}

\subsection{Enhanced Sonic Protocol with Batch Verification} \label{sec:enhanced}

By replacing the original restricted KZG polynomial commitment scheme by the one with batch verification, we present an enhanced version of zk-SNARK protocol (${\tt S}_{\tt EV}$) in Table \ref{tab:final_protocol}. To convert the interactive protocol ${\tt S}_{\tt EV}$ to be non-interactive, we can employ Fiat-Shamir heuristic to replace the verifier-supplied random challenges $(y, \beta, z)$ by hash values from the previous commitments.

\subsection{Implementation of Bushfire Insurance by Sonic Protocol} \label{sec:bushinsrspec}

In this section, we describe how to encode the bushfire detection model in Sec.\ref{sec:sensing} into a Sonic-compatible constraint system. We specify the detailed settings of Sonic-specific vectors $({\bf a}, {\bf b}, {\bf c})$ and $({\mathbf u_q, \mathbf  v_q, \mathbf  w_q}, k_q)$ in Tables~\ref{tab:linear_const} and \ref{tab:assignment}. We assume image size $n$ and $k$ is the length of bit indices.

We first construct the input vectors $({\bf a}, {\bf b}, {\bf c})$. Row $1$ to $n$ of these vectors are $\mathbf{n}^-$,  $(\mathbf r^- + \mathbf s^-)$ and $(\mathbf r^- - \mathbf s^- - \mathbf \theta^-)$ respectively. Because ${\bf a}\cdot{\bf b}={\bf c}$, it verifies $(\mathbf r^-_i - \mathbf s^-_i) = \mathbf{n}^-_i (\mathbf r^-_i + \mathbf s^-_i)+\mathbf \theta^-_i, i\in \{1,...,n\}$. Similarly, we construct row $n+1$ to $6n$ to verify the rest of the multiplication gates, which are shown in Table \ref{tab:assignment}. Row $5n+1$ to row $6n$ verifies $\bf i$ is binary and row $6n+1$ to $6n + (n + 2)k$ verifies $\bf e, \theta_d, G$ are a non-negative. Finally, row $6n + (n + 2)k + 1$ to the end is input from the data source. 

Next we construct $({\mathbf u_q, \mathbf  v_q, \mathbf  w_q}, k_q)$. The first step is to verify entries in $({\bf a}, {\bf b}, {\bf c})$ are consistent with the input from data sources. For example, row $1$ t $n$ of $\bf b$ are $\mathbf r^-+\mathbf s^-$, while $\mathbf r^-$ and $\mathbf s^-$ are row $6n + (n + 2)k + 1, ..., 8n + (n + 2)k$ of $\bf a$. Then the first constraint in Table \ref{tab:linear_const} checks $\bf a_{6n + (n + 2)k + i} + \bf a_{7n + (n + 2)k + i}=\bf b_{i}, i\in {1,...,n}$. The next step is to encode the sum gates in the circuits. For example, checking $\sum^n_{i=1} \mathbf i_i-G=\epsilon$, where $\bf i$ locates in row $5n+1$ to $6n$, and $G$ in row $6n + (n + 1)k + 1$ to $6n + (n + 2)k$ in binary decomposition form, and $\epsilon$ in $k_q$ as a public input. The last part checks binary decompositions are correct.

Proofs of binary integers and non-negative integers are based on the techniques in \cite{Flashproofs}. We refer the reader to the original paper for a detailed explanation. Here we illustrate these constraint systems $\mathscr C$ with two examples:

\begin{itemize}

\item
{\bf Example 1}: This problem is to decide if $w$ is binary: $w \in \{0, 1\}$. This example can be represented by a single constraint equation: $w (w-1) = 0$, equivalent to the following constraint system ${\mathscr C}$ with one multiplicative constraint and three linear constraints:
\[
\left\{
\begin{array}{r@{\ }l}
{\mathbf a}_1\cdot {\mathbf b}_1 & = {\mathbf c}_1 \\
{\mathbf a}_1 - {\mathbf b}_1 & = 0 \\
{\mathbf a}_1 - {\mathbf c}_1 & = 0 \\ 
{\mathbf a}_1 & = w 
\end{array}
\right. \qquad (\mathscr C)
\]
$({\mathbf a}_1, {\mathbf b}_1,  {\mathbf c}_1)$=$(w, w, w^2)$ is satisfiable in ${\mathscr C}$, if $w \in \{0, 1\}$.

\item
{\bf Example 2}: This problem is to decide if an integer $w$ is non-negative: $w \in \{0, ..., 2^{k} - 1\}$. This example can be represented by a set of constraint equations: 
\[
\left\{
\begin{array}{r@{\ }ll}
b_{i}  (b_{i} -2^{i-1}) & = 0, & \mbox{for\ }  1 \le i \le k \\
\sum_{i=1}^{k} b_i & = w &
\end{array}
\right.
\]
where $(b_i)_{i=1}^{k}$ represent the (scaled) bit-decomposition of $w$.
We can map the above set of constraint equations to a constraint system ${\mathscr C}$ with $k$ multiplicative constraints and $(2k+1)$ linear constraints:
\[
\left\{
\begin{array}{r@{\ }ll}
\mathbf a_i \cdot \mathbf b_i & = \mathbf c_i, & \mbox{for\ }  1 \le i \le k \\
{\mathbf a}_i - {\mathbf b}_i & = 0, & \mbox{for\ }  1 \le i \le k \\
{\mathbf a}_i 2^{i-1} - {\mathbf c}_i & = 0, & \mbox{for\ }  1 \le i \le k \\
\sum_{i=1}^{k} {\mathbf c}_i & = w
\end{array}
\right. \qquad (\mathscr C)
\]
$({\mathbf a}_i,{\mathbf b}_i,{\mathbf c}_i)$=$(b_{i}, b_{i}, b_{i}^2)$ is satisfiable in ${\mathscr C}$, if $w$ is non-negative and $(b_i)_{i=1}^{k}$ represent the (scaled) bit-decomposition of $w$. 
\end{itemize}

\begin{table*}[ht]
    \centering
    \caption{Settings of vectors $({\bf a}, {\bf b}, {\bf c})$ for parametric bushfire insurance.} \vspace{-10pt}
	{\scriptsize
    \begin{tabular}{c|c||@{}c|c|c@{}}
    \hline
    Row Index&Computation&{\bf a} & {\bf b} & {\bf c} \\
    \hline\hline
    
$1, ..., n$ & 
$(\mathbf r^-_i - \mathbf s^-_i) = \mathbf{n}^-_i (\mathbf r^-_i + \mathbf s^-_i)+\mathbf \theta^-_i, i\in \{1,...,n\}$ &
$(\mathbf n_i^-)_{i=1}^n$ & 
$(\mathbf r_i^-+\mathbf s_i^-)_{i=1}^n$  & 
$(\mathbf r_i^--\mathbf s_i^--\mathbf \theta_i^-)_{i=1}^n$\\

$n+1, ..., 2n$ & 
$(\mathbf r^+_i - \mathbf s^+_i) = \mathbf{n}^+_i (\mathbf r^+_i + \mathbf s^+_i)+\mathbf \theta^+_i, i\in \{1,...,n\}$ &
$(\mathbf n_i^+)_{i=1}^n$ & 
$(\mathbf r_i^++\mathbf s_i^+)_{i=1}^n$  & 
$(\mathbf r_i^+-\mathbf s_i^+-\mathbf \theta_i^+)_{i=1}^n$\\

$2n+1, ..., 3n$ &
$\mathbf{i}_i (\mathbf{n}^-_i - \mathbf{n}^+_i  - \kappa) = \sum_{j=1}^k \mathbf{e}_{i,j}, i\in \{1,...,n\}$&
$(\mathbf i_i)_{i=1}^n$ & 
$(\mathbf n_i^--\mathbf n_i^+-\kappa)_{i=1}^n$  &
$(\sum_{j=1}^k\mathbf e_{i,j})_{i=1}^n$\\

$3n+1, ..., 4n$ &
$(\mathbf \theta_i^-)^2, i\in \{1,...,n\}$&
$(\mathbf \theta_i^-)_{i=1}^n$ & 
$(\mathbf \theta_i^-)_{i=1}^n$  & 
${(\mathbf \theta_i^-)^2}_{i=1}^n$\\

$4n+1, ..., 5n$ & 
$(\mathbf \theta_i^+)^2, i\in \{1,...,n\}$&
$(\mathbf \theta_i^+)_{i=1}^n$ & 
$(\mathbf \theta_i^+)_{i=1}^n$  & 
${(\mathbf \theta_i^+)^2}_{i=1}^n$\\  

$5n+1, ..., 6n$ & 
$\mathbf i_i \in \{0,1\}, i\in \{1,...,n\}$&
$(\mathbf i_i)_{i=1}^n$ & 
$(1-\mathbf i_i)_{i=1}^n$  & 
$\mathbf 0$\\

$6n+1, ..., 6n+nk$ & 
$\mathbf e_i \ge 0$&
$(\mathbf e_{i,j})_{i=1, j=1}^{n,k}$ & 
$(\mathbf e_{i,j}-2^{j-1})_{i=1, j=1}^{n,k}$  & 
$\mathbf 0$\\

$6n+nk+1, ..., 6n+nk+k$ & 
$\theta_{d} \ge 0$&
$(\mathbf e_{\theta_{d},j})_{j=1}^{k}$ & 
$(\mathbf e_{\theta_{d},j}-2^{j-1})_{j=1}^{k}$  & 
$\mathbf 0$\\

$6n+(n+1)k+1, ..., 6n+(n+2)k$ & 
$G \ge 0$&
$(\mathbf e_{G,j})_{j=1}^{k}$ & 
$(\mathbf e_{G,j}-2^{j-1})_{j=1}^{k}$  & 
$\mathbf 0$\\

$6n+(n+2)k+1, ..., 7n+(n+2)k$ & Input from data source &$(\mathbf r_i^-)_{i=1}^n$ & $\mathbf 0$ & $\mathbf 0$\\
$7n+(n+2)k+1, ..., 8n+(n+2)k$ & Input from data source &$(\mathbf s_i^-)_{i=1}^n$ & $\mathbf 0$ & $\mathbf 0$\\
$8n+(n+2)k+1, ..., 9n+(n+2)k$ & Input from data source &$(\mathbf r_i^+)_{i=1}^n$ & $\mathbf 0$ & $\mathbf 0$\\
$9n+(n+2)k+1, ..., 10n+(n+2)k$ & Input from data source &$(\mathbf s_i^+)_{i=1}^n$ & $\mathbf 0$ & $\mathbf 0$\\

    \hline
    \end{tabular}
	}
    \label{tab:assignment}
\end{table*}

\subsection{Security of the Enhanced zk-SNARK Protocol}

There are several key properties of an argument system:

\begin{itemize}
    \item \textbf{Completeness}: For any $(x, w) \in {\mathbb F}_p^n \times {\mathbb F}_p^m$, such that $(x, w) \in {\mathscr R}_C$, $\textsf{\small Prove}$ should produce a valid proof $\pi$ to pass $\textsf{\small Verify}$. Namely, given $(x, w) \in {\mathscr R}_C$, if ${\tt srs} \leftarrow \textsf{\small KeyGen}(1^\lambda, C)$ and $\pi \leftarrow \textsf{\small Prove}({\tt srs},x,w)$, then $\textsf{\small Verify}({\tt srs},x,\pi) = {\tt True}$.

    \item \textbf{Knowledge Soundness}: Informally, a prover should not pass the verification, if the prover does not know the correct witness. Formally, for every successful polynomial-time adversary $\mathcal A$ who can provide a statement $x$ with a valid proof $\pi$, there exists a polynomial-time extractor ${\mathcal E}_{\mathcal A}$ who can extract the witness $w$ with high probability given the access to the adversary $\mathcal A$'s internal states:	
	\begin{align*}
		{\mathbb P}&\left [ \begin{array}{l}
            \textsf{\small Verify}({\tt srs},x,\pi) = {\tt True}  \\
            \wedge (x, w) \in {\mathscr R}_C
            \end{array}
            \middle|
            \begin{array}{l}
                {\tt srs} \leftarrow \textsf{\small KeyGen}(1^\lambda, C)\\
                (w, \pi) \leftarrow {\mathcal E}_{\mathcal A}({\tt srs},\pi)
            \end{array}
            \right] \\ &= 1 - \epsilon(\lambda) 
    \end{align*}

    \item \textbf{Perfect Honest-Verifier Zero Knowledge}: Any adversary $\mathcal A$ cannot distinguish between a valid proof $\pi$ of a statement and other random strings. Formally, for a honest verifier who faithfully follows the protocol, there exists a polynomial-time simulator $\textsf{\small Sim}$ who does not know the correct witness $w$, such that the distributions $D_1, D_2$ (defined below) are statistically indistinguishable:
    \begin{align*}
        D_1\triangleq&\{ {\tt srs} \leftarrow \textsf{\small KeyGen}(1^\lambda, C);\pi \leftarrow \textsf{\small Prove}({\tt srs},x,w) \} \\
        D_2\triangleq&\{ {\tt srs} \leftarrow \textsf{\small KeyGen}(1^\lambda, C); \pi \leftarrow \textsf{\small Sim}({\tt srs},x) \}
    \end{align*}

    \item \textbf{Succinctness}: $\textsf{\small Prove}({\tt srs}, x, w)$ generates proof $\pi$ with a very small size as compared with the size of public input $|x|$ or the witness $|w|$, and $\textsf{\small Verify}({\tt srs}, x, \pi)$ takes a very fast running time as compared with $|x|$. Sonic protocol has constant-sized proofs and constant verification time. The properties are preserved in the enhanced protocol.
	
    \item \textbf{Non-interactiveness}: There are several interactive protocols with {\em public coins}, meaning that the verifier must keep its internal state public. These interactive protocols with public coins can be converted to non-interactive protocols by {\em Fiat-Shamir heuristic} \cite{Fiat_Shamir_1987}. We presented the interactive zk-SNARK protocol with public coins in Table \ref{tab:datasource} and Table \ref{tab:final_protocol}, together with the modifications (using the Fiat-Shamir heuristic) to achieve non-interactive protocols in grey next to the interactive steps.
		
\end{itemize}

Next, we provide proofs of our enhanced zk-SNARK protocol for completeness and knowledge soundness and perfects honest-verifier zero knowledge.
\begin{enumerate}

    \item \textbf{Completeness} 
    Assume the restricted KZG is used as the commitment scheme. Given public input $\lambda, \hat{s}[X,Y], \hat{k}[Y], ({\tt pk}_j)_{j=1}^J$, and public data from $J$ data sources $d_j[X]$ (each of length $m_j$), The honest prover inputs $r[X,Y]$ and follows the protocol from step 1 to 7 correctly. As a result, the prover generates $6+J$ commitments $S_Y, K, R, \tilde{R}, T, S_X$, and $D_j, j\in \{1,...,J\}$, $8+J$ openings $r_1=r[z,1], r_2=r[zy, 1], t=t[z,y], \tilde{r}=\tilde{r}[z,1], k=\hat{k}[y], \hat{s}=\hat{s}[z, y], \hat{s_1}=\hat{s}[1, y], \hat{s_2}=\hat{s}[1, y], d_j=d_j[z], j\in \{1,...,J\}$ with their proofs, and $J$ signatures $\sigma_j$. 
    Then, in the last step, the verifier first verifies all the signatures are valid, which proves the commitments $D_j$ match those provided by the data sources. Then, it verifies all the openings are valid openings of corresponding commitments. Next, the verifier checks the correct computation of polynomials. First it checks $\hat{s_1}\stackrel{?}{=}\hat{s_2}=\hat{s}[1, y]$, which indicates the committed outsourced $\hat{s}[X,Y]$ matches the public input.
    Next it checks correct computation of $R[X,Y]$:
    \[
    r_1 = \tilde{r} + \sum_{j=1}^J d_j z^{N + \sum_{j=1}^{J-1} m _j}  
    \]
    It holds because of the homomorphism of the KZG commitment. 
    Finally, it checks 
    \[
    t \stackrel{?}{=} r_1(r_2+\hat{s})-k
    \] 
    This is verified by checking:
    \[
    t[z,y] \stackrel{?}{=} r[z,1](r[zy,1]+s[z,y])-\hat{k}[y]
    \]
    which holds by the definition of $t[X,Y]$. Therefore, the honest prover can pass all the verification and $\textsf{\small Verify}({\tt srs},x,\pi) = {\tt True}$.
    
    \item \textbf{Perfects Honest-Verifier Zero Knowledge} Assume an arbitrary polynomial-time simulator $\textsf{\small Sim}$ who can access all the public input of the protocol and the SRS strings from the $J$ data providers. It chooses random vectors $\mathbf a, \mathbf b$ from ${\mathbb F}_p$ of length $n$ and sets $\mathbf c = \mathbf a \cdot \mathbf b$. It then chooses $J$ random vectors $\mathbf d_1, ..., d_J$, of length $m_j$ for $j \in {1,...,J}$. Then the simulator computes $\tilde{r}[X,Y], d_1[X], ..., d_J[X]$ and $r[X,Y], t[X,Y]$. Then, the simulator performs the same as the prover described in Table \ref{tab:datasource} with respect to the polynomials. 
    As described in Section \ref{sec:dataprotocol}, the prover only reveals $5+2J$ evaluations of $r[X,Y]$, (including $r(z, 1), r(zy, 1), \tilde{r}, g^{r(X,1)}, g^{\tilde{r}(X,1)}$ and $d_j[z], g^{d_j}, j\in \{1,...,J\}$). Therefore, we set $6+2j$ random blinders with random coefficients and powers from $-2n-1$ to $-2n-6-2j$. Therefore, for a verifier obtained less than $6+2j$ openings, the prover's polynomials are indistinguishable from the random polynomials from the simulator. All other polynomials are either public input or computed from $r[X,Y]$, hence preserving perfect honest-verifier zero knowledge.

    \item \textbf{Knowledge Soundness}: We argue the knowledge soundness of the original Sonic protocol is preserved in the enhanced protocol. We made two modifications to original Sonic: (1) new batch verification of the restricted KZG and (2) validation of input sources. 
    
    First, we have proved the knowledge soundness of the new batch verification of restricted KZG in Theorem~\ref{thm:rKZGb}. Note that the restricted KGZ polynomial commitments are still used in the same manner as the original Sonic protocol, but only their verification can be batched more efficiently. Hence, as long as the batch verification is knowledge sound (an adversary cannot forge proof without knowing witnesses) under the Algebraic Group Model (AGM) - which is required by the proof of the original Sonic protocol, the soundness of Sonic protocol is still preserved by our new batch verification. 
    
    We also separate input data sources by validation of input sources. We apply signature schemes to validate the commitment of data from each source. Note that this is applied externally to Sonic protocol. Hence, this does not affect the security properties of Sonic protocol.

\end{enumerate}

\end{document}